\documentstyle[epsfig,prd,eqsecnum,aps,preprint,tighten]{revtex}

\begin{document}

\preprint{\vbox to 108 pt
{\hbox{IHES/P/98/54}\vfill}}
\draft

\title{Big Bang nucleosynthesis and tensor-scalar gravity}
\author{Thibault Damour}
\address{Institut des Hautes Etudes Scientifiques, 91440 
Bures-sur-Yvette, France \\
and \\
DARC, CNRS-Observatoire de Paris, 92195 Meudon, France}
\author{Bernard Pichon}
\address{DARC, CNRS-Observatoire de Paris, 92195 Meudon, France}

\maketitle

\begin{abstract}
Big Bang Nucleosynthesis (BBN) is studied within the framework of a 
two-parameter family of tensor-scalar theories of gravitation, with 
nonlinear scalar-matter coupling function $ a( \varphi) =  a_0 + \alpha_0 
( \varphi - \varphi_0) +  \frac{1}{2} \beta (\varphi - \varphi_0)^2$.
  We run a BBN code modified by tensor-scalar gravity,
 and  impose that the theoretically predicted BBN yields of Deuterium,
 Helium and Lithium lie within some conservative  observational ranges. 
It is found that large initial values of $a(\varphi)$
 (corresponding to cosmological expansion rates, for temperatures
higher than 1 MeV, much larger than standard)
 are compatible with observed BBN yields. However,
 the BBN-inferred upper bound on the cosmological baryon density is 
insignificantly modified  by considering tensor-scalar gravity.  Taking into 
account the effect of $e^+ e^-$ annihilation together with the subsequent 
effect of the matter-dominated era (which both tend to decouple $\varphi$ from 
matter), we find that the present value of the scalar coupling, i.e. the 
present level of deviation from Einstein's theory, must be, for compatibility
 with BBN, smaller than  $\alpha_0^2 \lesssim 10^{-6.5} \, \beta^{-1} \,
(\Omega_{\rm matter} \, h^2 / 0.15)^{-3/2}$ when $\beta \gtrsim 0.5$~.
\end{abstract}

\newpage

\section{Introduction}

The idea that Einstein's tensor gravitational field might be accompanied by 
a massless scalar partner was first suggested in the twenties by Kaluza 
\cite{K21}. Since then, such tensor-scalar gravity theories have been 
studied in detail by many authors \cite{J49,F56,BD61,N70,W70,DEF92}. String 
theory has recently revived 
the motivation for considering gravitational-strength scalar fields, such 
as the (model-independent) dilaton or the (Kaluza-Klein-type) moduli (see, 
e.g., \cite{GSW}). In the simplest versions of tensor-scalar theories of 
gravity (those respecting the equivalence principle) the coupling between 
matter and the scalar field $\varphi$ is described by a single ``coupling 
function'' $a(\varphi)$, such that all physical mass scales get multiplied 
by a factor $A(\varphi) \equiv \exp [a(\varphi)]$ when measured in 
``Einstein units'' (see below for the definition of the Einstein conformal 
frame). For instance, the Jordan-Fierz-Brans-Dicke (JFBD) theory is the 
one-parameter theory defined by a linear coupling function, $a(\varphi) = 
\alpha_0 \, \varphi$. The JFBD theory is not an appealing alternative to 
general relativity because its only (dimensionless) parameter $\alpha_0$ 
needs to be fine-tuned to a small value, $\alpha_0^2 < 10^{-3}$, to be 
consistent with existing experimental data. [See, e.g., \cite{DEF} for a 
discussion of the constraints on tensor-scalar gravity brought by 
solar-system, and binary pulsar, data.] By contrast, it has been shown that 
more general theories with nonlinear coupling functions, containing no small 
parameters, can be naturally compatible with experimental data because the 
cosmological evolution drives the background value of $\varphi$ toward a 
value that minimizes the coupling function $a(\varphi)$, thereby reducing 
by a large factor all the present observable effects of $\varphi$ 
\cite{DN}, \cite{DP}. Within such cosmological-attractor models, the only 
regime in which the scalar field can play a quantitatively important role 
is early cosmology.

Big Bang nucleosynthesis (BBN) is the earliest cosmological process that is 
physically well established and about which one has reasonably accurate 
observational data. Its importance for constraining many physical or 
astrophysical scenarios was first pointed out by Shvartsman \cite{S69}. For 
recent treatments see, e.g., \cite{Reeves,CST95,lmk}. In particular, 
BBN is crucially used for deriving an upper bound on the cosmological 
baryon density $\Omega_b = \rho_{\rm baryon} / \rho_{\rm closure}$ which, 
even for extreme parameters, is claimed to be $\Omega_b < 0.03 \, h^{-2} < 
0.2$ \cite{Reeves,CST95,lmk} (here, $h \equiv H_0 / (100 \, {\rm km \, 
s}^{-1} \ {\rm Mpc}^{-1})$). One of the main motivations of the present 
paper is to examine whether the presence of a gravitational-strength scalar 
field $\varphi$, having a generic non-JFBD, i.e., nonlinear, coupling 
function $a(\varphi)$, can significantly modify the standard limit on 
$\Omega_b$, while being still compatible with present gravitational 
experiments. A secondary motivation is to study the level of present scalar 
admixture to Einstein's gravity (observable, in principle, in precision 
tests of relativistic gravity) which is naturally compatible with BBN data, 
i.e. with observed light element abundances. 

Our analysis is more general 
and/or more exact than previous attacks on this problem. Indeed, we 
consider the case of a nonlinear coupling function $a(\varphi)$ admitting a 
local minimum, while most previous studies restricted themselves to 
JFBD-type theories \cite{RM82,DG91,CGQ92} (see also the 
related studies of BBN limits on the variability of the gravitational 
constant $G$ \cite{S69}, \cite{S76,B78,YSSR79,AKR90}). A recent work by 
Santiago et al. \cite{SKW} has considered 
nonlinear coupling functions $a(\varphi)$ of the type we shall study but 
their analysis is, in our opinion, less satisfactory than ours
  in two\footnote{In addition, the solution for 
$\varphi$ during the end of the radiation-dominated epoch is incorrectly 
given in Sec.~IV~D of Ref.~\cite{SKW}: e.g., their Eq.~(4.37) should 
contain $J_1 (x) / x$ and $Y_1 (x) / x$.} ways: (i) they approximated the 
effect of $\varphi$ as a {\it small, constant speed up factor} $\xi_n$ 
affecting only the value of the neutron/proton ratio at 
some effective ``freeze-out'', supposed to take place before and
separately from $e^+ e^-$ annihilation, without 
recomputing in tensor-scalar gravity the production of all the light 
elements, and (ii) they related the value of $\varphi$ (and the 
corresponding $\xi_n$) at freeze-out to its present value $\varphi_0$ by 
{\it integrating backward in time}. Concerning the point i), we shall, 
instead, find that the beginning of $e^+ e^-$ annihilation generates a
source term for the evolution of $\varphi$ which generically (especially
for largish values of $\beta$) cannot be
neglected during freeze-out so that the speed up factor can vary a lot
during the critical freezing of the $n/p$ ratio. ( see Section 4 for
a more detailed discussion).  Concerning ii), the 
procedure of backward time-integration  may generically lead to incorrect 
results because the $\varphi$ evolution equation is similar to that of a 
damped harmonic oscillator \cite{DN}. Integrating backward in time such an 
equation is equivalent to integrating forward in time an oscillator with 
negative friction, which introduces spurious runaway solutions.
The origin of the introduction of these runaway solutions is simply
that, when integrating backward in time, one lacks the correct 
``initial'' condition for the present time derivative of $\varphi$,
$\dot{\varphi}_0$, corresponding to some $\varphi_0$. [Indeed,
$\dot{\varphi}_0$ is nonlocally defined by the full forward 
(damped) evolution of $\varphi$ starting from the physically natural
initial condition, $\dot{\varphi}_{\rm in} = 0$, deep into the radiation
era (see below).] We suppose that this 
pollution by runaway solutions is the root of the (incorrect)
obtention of infinite peaks in the figures of Ref.~\cite{SKW}, 
corresponding to speed-up factors equal to one (which is 
physically forbidden, as we shall discuss below).
 [We also suppose that this pollution by runaway solutions,
made more critical because of a direct numerical integration
from the present time back to BBN time,
invalidates the physical relevance of the work of Ref.~\cite{SA}, 
 and explains why their limits on the coupling parameter
$\alpha_0^2$ are about fourteen orders of magnitude smaller than the
limits we get here.] However, we have not checked our supposition
(concerning the effect of runaway solutions) by running reverse
integrations ourselves.

In this work, we shall avoid this 
potential problem of unphysical runaway
 solutions by integrating only forward in 
time. Another difference with previous work is that we shall compute, using 
a full BBN code modified by tensor-scalar gravity effects, the abundances 
of all the light elements: Deuterium, Helium 3, Helium 4 and Lithium 7. 
Indeed, though they are still large uncertainties on the primordial 
abundances of these elements, it is important to combine the predicted 
abundances of all the light elements with the corresponding observational 
bounds to get consistent constraints on tensor-scalar gravity.

 The present paper works within the framework defined by the following
assumptions: (i) we consider tensor-scalar gravitation theories 
containing a single massless scalar field $\varphi$, (ii) the coupling
function of $\varphi$ is restricted to the two-parameter family 
$ a( \varphi) =  a_0 + \alpha_0
( \varphi - \varphi_0) +  \frac{1}{2} \beta (\varphi - \varphi_0)^2$
(see Section II for  the choice of this quadratic form), (iii) we
require that deep into the radiation era the time derivative of 
$\varphi$ vanishes (see our detailed discussion below), and (iv)
we consider the case of a spatially flat Friedmann universe (the effect
of non zero curvature has been investigated in Ref. \cite{DN}).

\section{Tensor-scalar gravity theories}

We consider tensor-scalar gravitation theories containing a single 
(massless) scalar field, assumed to couple to the trace of the 
energy-momentum tensor. This coupling means that the scalar source is (for 
bodies having negligible self gravity) proportional to the (inertial) mass, 
so that the equivalence principle is respected. The most general theory 
describing such a mass-coupled long-range scalar field contains one arbitrary 
coupling function \cite{F56}
\begin{equation}
A (\varphi) \equiv \exp (a (\varphi)) \, . \label{eq2.1}
\end{equation}
The action defining the theory reads
\begin{equation}
S = \frac{1}{16 \pi G_*} \int d^4 x \, g_*^{1/2} (R_* - 2 g_*^{\mu \nu} 
\partial_{\mu} \varphi \, \partial_{\nu} \varphi) + S_m [\psi_m ; A^2 
(\varphi) g_{\mu \nu}^*] \, . \label{eq2.2}
\end{equation}
Here, $G_*$ denotes a bare gravitational coupling constant, $R_* \equiv 
g_*^{\mu \nu} \, R_{\mu \nu}^*$ the curvature scalar of the ``Einstein 
metric'' $g_{\mu \nu}^*$ describing the pure spin-2 excitations, and 
$\varphi$ the long-range scalar field describing spin-0 excitations. [We 
use the signature $-+++$ and the notation $g_* \equiv - \det g_{\mu 
\nu}^*$.] The last term in Eq.~(\ref{eq2.2}) denotes the action of matter, 
which is a functional of some matter variables (collectively denoted by 
$\psi_m$) and of the (Jordan-Fierz) ``physical metric''
\begin{equation}
\widetilde{g}_{\mu \nu} \equiv A^2 (\varphi) \, g_{\mu \nu}^* \, . 
\label{eq2.3}
\end{equation}
Laboratory clocks and rods measure the metric $\widetilde{g}_{\mu \nu}$ 
which, in the model considered here, is universally coupled to matter. In 
the BBN context, the standard laws of nongravitational physics (such as 
nuclear reaction rates and thermodynamical laws) will hold in their usual 
form when expressed in ``physical units'', i.e. in units of the proper 
interval $d \widetilde{s}^2 = \widetilde{g}_{\mu \nu} \, dx^{\mu} \, 
dx^{\nu}$. For instance, N\oe ther's theorem applied to the matter action 
$S_m [\psi_m ; \widetilde{g}_{\mu \nu}]$ yields the usual law of 
conservation of energy and momentum:
\begin{equation}
\widetilde{\nabla}_{\nu} \, \widetilde{T}^{\mu \nu} = 0 \, , \label{eq2.4}
\end{equation}
where $\widetilde{T}^{\mu \nu} \equiv 2 \, \widetilde{g}^{-1/2} \, \delta 
S_m / \delta \, \widetilde{g}_{\mu \nu}$ is the material stress-energy 
tensor in physical units. In Eq.~(\ref{eq2.4}) the covariant derivative 
$\widetilde{\nabla}_{\nu}$ is that defined by $\widetilde{g}_{\mu \nu}$.

On the other hand, the gravitational field equations of the theory are most 
simply formulated in terms of the pure-spin variables $(g_{\mu \nu}^* , 
\varphi)$. From (\ref{eq2.2}), they read
\begin{mathletters}
\label{eq2.5}
\begin{eqnarray}
&&R_{\mu \nu}^* = 2 \, \partial_{\mu} \varphi \, \partial_{\nu} \varphi + 8 
\pi \, G_* \left( T_{\mu \nu}^* - \frac{1}{2} \, T^* \, g_{\mu \nu}^* 
\right) \, , \label{eq2.5a} \\
&&\Box_{g^*} \, \varphi = - 4\pi \, G_* \, \alpha (\varphi) \, T_* \, , 
\label{eq2.5b}
\end{eqnarray}
\end{mathletters}
with $T_*^{\mu \nu} \equiv 2 \, g_*^{-1/2} \, \delta S_m / \delta \, g_{\mu 
\nu}^*$ denoting the material stress-energy tensor in Einstein units. It is 
related to the physical-units stress-energy tensor through
\begin{equation}
g_{\nu \sigma}^* \, T_*^{\mu \sigma} \equiv T_{*\nu}^{\mu} = A^4 (\varphi) 
\, \widetilde{T}^{\mu}_{\nu} \equiv A^4 (\varphi) \, \widetilde{g}_{\nu 
\sigma} \, \widetilde{T}^{\mu \sigma} \, . \label{eq2.6}
\end{equation}
The quantity $\alpha (\varphi)$ on the R.H.S. of Eq.~(\ref{eq2.5b}) plays a 
crucial role in the theory. It is the logarithmic derivative of the 
coupling function,
\begin{equation}
\alpha (\varphi) \equiv \frac{\partial \, \ln \, A(\varphi)}{\partial \, 
\varphi} \equiv \frac{\partial \, a(\varphi)}{\partial \, \varphi} \, , 
\label{eq2.7}
\end{equation}
and it measures the basic (field-dependent) coupling strength between the 
scalar field and matter. The JFBD theory is defined by a linear field 
dependence of $a(\varphi) = \ln \, A(\varphi) = \alpha_0 \, \varphi$, i.e. 
by a constant (field-independent) coupling strength $\alpha (\varphi) = 
\alpha_0$ [$\alpha_0^2 = (2 \omega + 3)^{-1}$ in the notation of 
\cite{BD61}]. Generically, one might expect a nonlinear field dependence of 
$a(\varphi)$ leading to a field-dependent coupling strength $\alpha 
(\varphi)$. It has been shown in Refs.~\cite{DEF92}, \cite{DEF96a} that all 
{\it weak field} (``post-Newtonian'') deviations from general relativity 
(of any post-Newtonian order) can be expressed in terms of the values of 
$\alpha (\varphi)$ and of its successive $\varphi$-derivatives, starting 
with
\begin{equation}
\beta (\varphi) \equiv \frac{\partial \, \alpha (\varphi)}{\partial \, 
\varphi} \, , \label{eq2.8}
\end{equation}
at the present ``vacuum expectation value'' $\varphi_0$ of the field 
$\varphi$. Here $\varphi_0$ denotes the asymptotic value of $\varphi$ at 
spatial infinity, at the present epoch. At the first post-Newtonian 
approximation, deviations from general relativity are proportional to the 
well-known Eddington-Nordtvedt-Will parameters
\begin{mathletters}
\label{eq2.9}
\begin{eqnarray}
&&\overline{\gamma} \equiv \gamma_{\rm Edd} - 1 = -2 \, \alpha_0^2 / 
(1+\alpha_0^2) \, , \label{eq2.9a} \\
&&\overline{\beta} \equiv \beta_{\rm Edd} - 1 = + \frac{1}{2} \, \beta_0 \, 
\alpha_0^2 / (1+\alpha_0^2)^2 \, , \label{eq2.9b}
\end{eqnarray}
\end{mathletters}
where $\alpha_0 \equiv \alpha (\varphi_0)$ and $\beta_0 \equiv \beta 
(\varphi_0)$. We see explicitly from Eqs.~(\ref{eq2.9}) that post-Newtonian 
deviations from general relativity tend to zero with $\alpha_0$ at least as 
fast as $\alpha_0^2$. This holds true for {\it weak-field} deviations of 
arbitrary post-Newtonian order \cite{DEF96a}. By contrast, it was found in 
Ref.~\cite{DEFnonpert} that {\it strong-field} deviations from general 
relativity do not tend to zero with $\alpha_0$ if the parameter $\beta_0 
\equiv \beta (\varphi_0) \equiv \partial \, \alpha (\varphi_0) / \partial 
\, \varphi_0$ is sufficiently negative. However, if one considers (as we 
shall do here) the case where the cosmological-attractor mechanism of 
Refs.~\cite{DN}, \cite{DP} takes place, the parameter $\beta_0$ is 
necessarily positive. Indeed, Refs.~\cite{DN}, \cite{DP} found that the 
spatial average of $\varphi$ is attracted, during the cosmological 
evolution, toward a {\it minimum} of the coupling function $a(\varphi)$. 
Therefore, the present cosmological value of $\varphi$, $\varphi_0 = 
\varphi (t_0)$, is generically expected to be very close to a value 
$\varphi_m$ such that $\alpha_m = \partial \, a(\varphi_m) / \partial \, 
\varphi_m = 0$ and $\beta_m = \partial^2 \, a(\varphi_m) / \partial \, 
\varphi_m^2 > 0$. In such a case, all present deviations from general 
relativity (weak-field ones and strong-field ones alike) are expected to be 
very small, because they all contain a factor $\alpha_0^2 \simeq [\beta_m 
(\varphi_0 - \varphi_m)]^2 \ll 1$. 

As was discussed in Refs.~\cite{DN}, \cite{DP}, and is recalled in Section
3 below, the Einstein-time evolution for $\varphi$ deep into the radiation
dominated era can be approximated by 
$\ddot{\varphi} + 3 H_* \, \dot{\varphi} \simeq 0$, which shows that 
$\dot{\varphi}$ decreases, during the expansion, as the inverse cube of the
Einstein-frame scale factor $R_*$ (in usual cosmological parlance, one 
can say that $\dot{\varphi}$ contains only a decreasing mode).
 Therefore the  natural initial conditions for
$\varphi$ deep into the radiation era  are that $\dot{\varphi}$ vanishes,
while $\varphi = \varphi_{\rm in}$ has an arbitrary value. Starting from
these values, $\varphi$ will run {\it down} the coupling function 
$a(\varphi)$ and will be attracted to the {\it nearest} minimum of 
$a(\varphi)$. [Note that this shows, as emphasized in Ref.~\cite{DN},
that the value of $a(\varphi)$ irreversibly decreases during the expansion,
so that the speed up factor must always be larger than one.]
As the generic behaviour of a function 
near a minimum is parabolic, $a(\varphi) = a(\varphi_m) +
 \frac{1}{2} \, \beta_m (\varphi 
- \varphi_m)^2 + {\cal O} ((\varphi - \varphi_m)^3)$, it is plausible to 
assume that  $a(\varphi)$ is roughly parabolic in the whole interval around
$\varphi_m$ containing $\varphi_{\rm in}$. We shall therefore follow 
Ref.~\cite{DEF} and study the paradigmatic case where the coupling function 
is quadratic in $\varphi$, $a_{\rm quad} (\varphi) = a_0 + \alpha_0 
(\varphi - \varphi_0) + \frac{1}{2} \, \beta_0 (\varphi - \varphi_0)^2$. 
This case is the simplest generalization of the JFBD theory ($a_{\rm JFBD} 
(\varphi) = a_0 + \alpha_0 (\varphi - \varphi_0)$) admitting a nonlinear 
coupling function, and susceptible of being cosmologically attracted to a 
point where $\varphi$ decouples from matter. By appropriately choosing the 
unit of length one can fix $a_0$ to zero. One can also conventionally fix 
the minimum of $a(\varphi)$ at $\varphi_m \equiv 0$. Therefore, without 
loss of generality, we shall write the general ``quadratic model'' in the 
simple form
\begin{equation}
a_{\rm quad} (\varphi) \equiv \ln \, A_{\rm quad} (\varphi) = \frac{1}{2} 
\, \beta \, \varphi^2 \, , \label{eq2.10}
\end{equation}
with $\beta > 0$. This model contains two free parameters: the quantity 
$\beta = \partial^2 \, a (\varphi) / \partial \, \varphi^2$ ($=\beta_0 = 
\beta_m$), and, either the initial value $\varphi_{\rm in}$ of $\varphi$ deep 
in 
the radiation dominated era, or, alternatively, the present value 
$\varphi_0$ of $\varphi$. The aim of the present paper is to investigate 
which values of $\beta$ and $\varphi_{\rm in}$ (and consequently, 
$\varphi_0$) 
are compatible with BBN data.

\section{Big Bang nucleosynthesis in tensor-scalar gravity}

As mentioned in Section 2, the standard laws of nongravitational physics 
will keep their usual form if one measures lengths and times in ``physical 
units'', i.e. in units of the interval $d \widetilde{s}^2 = 
\widetilde{g}_{\mu \nu} \, dx^{\mu} \, dx^{\nu} = A^2 (\varphi) \, ds_*^2$. 
In a cosmological context, the physical interval will read
\begin{equation}
d \widetilde{s}^2 = - d \widetilde{t}^2 + \widetilde{R}^2 (\widetilde{t}) 
\, d\ell^2 \, , \label{eq3.1}
\end{equation}
where $\widetilde{t}$ is the physical cosmological time, and 
$\widetilde{R}$ the physical scale factor. Here
\begin{equation}
d\ell^2 = \frac{dr^2}{1-kr^2} + r^2 (d\theta^2 + \sin^2 \theta \, d 
\phi^2) \, , \label{eq3.2}
\end{equation}
is the metric of a 3-space of constant curvature $k = +1$, $0$ or $-1$. The 
corresponding physical expansion rate (Hubble parameter) is
\begin{equation}
\widetilde H = \frac{d}{d \widetilde t} \, \ln \, \widetilde R \, . 
\label{eq3.3}
\end{equation}

The Einstein-frame counterparts of the above quantities are
\begin{equation}
ds_*^2 = - dt_*^2 + R_*^2 (t_*) \, d\ell^2 \, , \label{eq3.4}
\end{equation}
\begin{equation}
H_* = \frac{d}{dt_*} \, \ln \, R_* \, . \label{eq3.5}
\end{equation}
The relation $d\widetilde{s}^2 = A^2 (\varphi) \, ds_*^2$ gives the links
\begin{mathletters}
\label{eq3.6}
\begin{eqnarray}
&&d \widetilde t = A(\varphi) \, dt_* \, , \label{eq3.6a} \\
&&\widetilde R = A(\varphi) \, R_* \, . \label{eq3.6b}
\end{eqnarray}
\end{mathletters}
Let us introduce the notation $\chi$ for the Einstein-time derivative of 
$\varphi$:
\begin{equation}
\chi \equiv \frac{d\varphi}{dt_*} \equiv A(\varphi) \, \frac{d\varphi}{d 
\widetilde t} \, . \label{eq3.7}
\end{equation}
In terms of this notation, the links (\ref{eq3.6}) give the following 
relation between the Hubble parameters,
\begin{equation}
\widetilde H = (A(\varphi))^{-1} \, [H_* + \alpha (\varphi) \, \chi ] \, , 
\label{eq3.8}
\end{equation}
where $\alpha (\varphi) \equiv \partial \, a (\varphi) / \partial \, 
\varphi$.

We computed BBN by using Kawano's update \cite{kawano} of Wagoner's code 
\cite{wagoner}. This code evolves thermodynamical variables and nuclear 
abundances as functions of the physical temperature $\widetilde T$. All 
these physical evolution equations are unchanged in tensor-scalar gravity. 
The only modifications that need to be brought to this code are: (i) a 
modified expression for the physical expansion rate $\widetilde H$ as a 
function of the energy density, and (ii) two new first-order equations 
giving the evolution of the scalar field. Note that we assume three light 
neutrinos throughout this work.

To get the modified expression for the expansion rate we need to write 
explicitly the modified Einstein equations (\ref{eq2.5a}). In the present 
cosmological context they yield
\begin{mathletters}
\label{eq3.9}
\begin{eqnarray}
&&- \frac{3}{R_*} \, \frac{d^2 \, R_*}{dt_*^2} = 4\pi \, G_* (\rho_* + 3 
p_*) + 2 \, \chi^2 \, , \label{eq3.9a} \\
&&3 \, H_*^2 + 3 \, \frac{k}{R_*^2} = 8\pi \, G_* \, \rho_* + \chi^2 \, . 
\label{eq3.9b}
\end{eqnarray}
\end{mathletters}
Here $\rho_*$ and $p_*$ are the Einstein-units energy density and pressure, 
linked to their physical counterparts by Eq.~(\ref{eq2.6}), i.e. by
\begin{equation}
\rho_* = A^4 \, \widetilde{\rho} \ , \quad p_* = A^4 \, \widetilde p \, , 
\label{eq3.10}
\end{equation}
where $A \equiv A (\varphi)$. As usual, the curvature term $k / R_*^2$ in 
Eq.~(\ref{eq3.9b}) is negligible during BBN so that Eq.~(\ref{eq3.9b}) gives 
the following result for $H_*$ in terms of the physical energy density and 
$\chi \equiv d \varphi / dt_*$
\begin{equation}
H_* = \sqrt{\frac{8\pi \, G_*}{3} \, A^4 \, \widetilde{\rho} + \frac{1}{3} 
\, \chi^2} \, . \label{eq3.11}
\end{equation}
The corresponding physical Hubble parameter is then computed by using 
Eq.~(\ref{eq3.8}).

It remains to write explicit evolution equations for the scalar field 
$\varphi$. Eq.~(\ref{eq2.5b}) yields
\begin{equation}
\frac{d^2 \varphi}{dt_*^2} + 3 \, H_* \, \frac{d\varphi}{dt_*} = -4\pi \, 
G_* \, \alpha (\varphi) \, (\rho_* - 3 p_*) \, . \label{eq3.12}
\end{equation}
In terms of the variable $\chi$, Eq.~(\ref{eq3.7}), and of the physical 
time $\widetilde t$, this gives the first-order system
\begin{mathletters}
\label{eq3.13}
\begin{eqnarray}
&&\frac{d\varphi}{d \widetilde t} = A^{-1} \, \chi \, , \label{eq3.13a} \\
&&\frac{d\chi}{d \widetilde t} = - A^{-1} \, [3 \, H_* \, \chi + 4\pi \, 
G_* \, \alpha (\varphi) \, A^4 \, \widetilde{\sigma}] \, , \label{eq3.13b}
\end{eqnarray}
\end{mathletters}
where $\widetilde{\sigma}$ denotes the following ``source term''
\begin{equation}
\widetilde{\sigma} = \widetilde{\rho} - 3 \widetilde p \, . \label{eq3.14}
\end{equation}
Around the period of primordial nucleosynthesis, the scalar source term can 
be approximated by
\begin{equation}
\widetilde{\sigma}_{\rm BBN}
 \simeq \widetilde{\sigma}_e \equiv \widetilde{\rho}_e - 
3 \widetilde{p}_e \, , \label{eq3.15}
\end{equation}
where the index ``$e$'' denotes the contribution from electrons and 
positrons. Indeed, among the various contributions to the total energy 
entering Eq.~(\ref{eq3.11}),
\begin{equation}
\widetilde{\rho} = \widetilde{\rho}_{\gamma} + \widetilde{\rho}_e + 
\widetilde{\rho}_{\nu} + \widetilde{\rho}_m \, , \label{eq3.16}
\end{equation}
the massless (or, at least, ultrarelativistic) photon and neutrino 
contributions $\widetilde{\rho}_{\gamma}$, $\widetilde{\rho}_{\nu}$ satisfy 
$\widetilde{\sigma}_{\gamma} = 0 = \widetilde{\sigma}_{\nu}$, while the non 
relativistic matter contribution $\widetilde{\rho}_m$ (cold dark matter 
plus baryons) is negligibly small (it becomes, however, important later, 
around matter-radiation equivalence and during the matter dominated era). By 
contrast, it 
happens by coincidence that one of the most important phenomena of BBN, 
the freezing out of the neutron to proton ratio, takes place around a 
physical temperature $\widetilde{T}_F \sim 1 \, {\rm Me V}$, which is 
precisely when the $e^+ \, e^-$ plasma starts becoming non relativistic 
before annihilating. In other words, the freeze-out of $(n/p)$ takes place 
when there is a numerically large contribution $\widetilde{\sigma}_e$ to 
the source driving the evolution of the scalar field $\varphi$. This 
fact has been taken into account only in the coarse 
approximation of considering that the freeze-out takes place before, and 
separately from, $e^+ \, e^-$ annihilation  in Ref.~\cite{SKW}.
 Our full numerical 
integration of a scalar-tensor BBN code shows that $\widetilde{\sigma}_e$ 
drives a strong (generically oscillatory)
 evolution of $\varphi$ at the same time (if not before) the 
$(n/p)$ ratio freezes out (See Section IV C below).

Summarizing, we completed an existing standard BBN code \cite{wagoner}, 
\cite{kawano} by adding: (i) Eqs.~(\ref{eq3.8}), (\ref{eq3.11}) for 
computing $\widetilde H$ in terms of $\widetilde{\rho}$, $\varphi$ and 
$\chi$, and (ii) Eqs.~(\ref{eq3.13}) for computing the $\widetilde 
t$-evolution of $\varphi$ and $\chi$. The source terms in these additional 
equations were taken to be (\ref{eq3.15}) and (\ref{eq3.16}). We used for 
the crucial scalar source $\widetilde{\sigma}_e$ the expansion
\begin{equation}
\widetilde{\sigma}_e = \frac{2}{\pi^2} \, \widetilde{m}_e^4 \,  
\sum_{n=1}^{\infty} \, (-)^{n+1} \, \frac{K_1 (nz)}{nz} \, . 
\label{eq3.16+}
\end{equation}
Here, $K_1$ is a modified Bessel function, and $z \equiv \widetilde{m}_e / 
\widetilde T$ (in units where $\hbar = c =k = 1$). As this is an alternating
series, it is numerically important to retain an {\it even} (and large enough)
number of terms. We kept sixteen terms in the expansion (\ref{eq3.16+}). [ By
contrast, Kawano's code uses five terms. Note also that the coefficient in 
front of this expansion is misprinted\footnote{ We were informed by the authors
of Ref.~\cite{SKW}, that they used the correct value in their code.}
 in Ref.~\cite{SKW}.] See 
Eq.~(\ref{eq3.22}) below for the exact, unexpanded expression of 
$\widetilde{\sigma}_e$.

The rest of the code computes the evolution of thermodynamical variables 
and of nuclear abundances in terms of the physical temperature $\widetilde 
T$. The link between physical temperature $\widetilde T$ and physical time 
$\widetilde t$ is given by the standard thermodynamical relation
\begin{equation}
\frac{d \widetilde T}{d \widetilde t} = -3 \, \frac{\widetilde{\rho} +
\widetilde p}{d \widetilde{\rho} / d \widetilde T} \, \widetilde H \, . 
\label{eq3.17}
\end{equation}
The input data for each run of the code are:
\begin{equation}
\eta = (4/11) \, (\widetilde{n}_b / \widetilde{n}_{\gamma})_{\rm in} \ , \ 
\varphi_{\rm in} \ , \ \chi_{\rm in} \ \hbox{and} \ \beta \, . 
\label{eq3.18}
\end{equation}
The coefficient $4/11$ is introduced so that $\eta$ measures the final 
baryon-to-photon ratio, $\eta = (\widetilde{n}_b / 
\widetilde{n}_{\gamma})_{\rm out}$, obtained after $e^+ e^-$ annihilation. As 
is discussed below, when $\beta$ is larger than about 0.2, the bare 
Newton constant $G_*$ entering Eqs.~(\ref{eq3.11}) and (\ref{eq3.13b}) can 
be taken to be numerically equal to the value presently measured in 
Cavendish experiments, $G_* \simeq 6.672 \times 10^{-8}$ cm$^3$ ${\rm g}^{-1} 
\, {\rm s}^{-2}$. [This is because the cosmological attraction of $\varphi$ 
toward 
$\varphi_m = 0$ is quite efficient when $\beta \geq 0.2$, so that the 
presently observable value $\widetilde G (\varphi_0) = G_* \, A^2 
(\varphi_0) \, [1+\alpha^2 (\varphi_0)]$ differs negligibly from $G_*$.] 
The output data for each run are then\footnote{We do not mention $^7{\rm 
Be}$ and $^8{\rm Li}$ which we shall not use.}
\begin{equation}
\varphi_{\rm out} \ , \ {\rm D} / {\rm H} \ , \ ^3{\rm He} / {\rm H} \ , \ Y 
(^4{\rm He}) \ \hbox{and} \ ^7{\rm Li}/{\rm H} \, . \label{eq3.19}
\end{equation}
The ratio ${\rm D} / {\rm H} \equiv n({\rm D}) / n({\rm H})$ denotes, for 
instance, the number of 
Deuterium nuclei per proton after BBN, while $Y (^4{\rm He}) \simeq 4n 
(^4{\rm He}) / (n({\rm H}) + 4n (^4{\rm He}))$ is the abundance by weight of 
Helium 4. The other physical parameters needed in the code are taken 
to have their currently measured values (in physical units). In particular, 
the neutron (exponential) lifetime is taken to be $\tau_n = 888.54 \, {\rm 
s}$.

An important issue (which, in our opinion, was  not properly
dealt with in previous work) 
concerns the initial values for $\varphi$ and its derivative $\chi = d 
\varphi / dt_*$. Our numerical calculations integrate the evolution for 
$\varphi$ in the normal direction, i.e. forward in time. If we insert our 
paradigmatic quadratic coupling function (\ref{eq2.10}), leading to the 
simple linear coupling strength
\begin{equation}
\alpha_{\rm quad} (\varphi) = \beta \, \varphi \, , \label{eq3.20}
\end{equation}
into Eq.~(\ref{eq3.12}), we see that $\varphi (t_*)$ satisfies the equation 
(here $\dot{\varphi} \equiv d \varphi / dt_*$, $\sigma_* \equiv \rho_* - 3 
p_*$)
\begin{equation}
\ddot{\varphi} + 3 H_* \, \dot{\varphi} + 4\pi \, G_* \, \sigma_* \, \beta 
\, \varphi = 0 \, . \label{eq3.21}
\end{equation}
This is the equation of a {\it damped} harmonic oscillator with 
time-varying {\it positive friction} $f=3H_*$, and time-varying positive 
spring constant $\omega_0^2 (t_*) \equiv 4\pi \, \beta \, G_* \, \sigma_*$. 
Indeed, the scalar source $\sigma_* = \rho_* - 3 p_* = A^4 
(\widetilde{\rho} - 3 \widetilde p)$ is always positive for a thermal 
distribution of particles \cite{DN}, \cite{DP},
\begin{equation}
\widetilde{\rho}_A - 3 \widetilde{p}_A = \frac{g_A \, 
\widetilde{m}_A^2}{2\pi^2} \int_0^{\infty} \frac{dq \, q^2}{\widetilde{E}_A 
[\exp (\widetilde{E}_A / \widetilde T) \pm 1]} \ , \ \widetilde{E}_A = 
\sqrt{\widetilde{m}_A^2 + q^2} \, . \label{eq3.22}
\end{equation}
Here, $\widetilde{m}_A$ is the mass of the considered particle, $g_A$ is 
its number of degrees of freedom, and the upper (lower) sign holds for 
Fermions (Bosons). As emphasized in Refs.~\cite{DN}, \cite{DP} the positive 
damping in Eq.~(\ref{eq3.21}) means that when one starts the $\varphi$ 
evolution very early in the radiation-dominated era (where the total number 
of effective relativistic degrees of freedom is large) and/or for 
temperatures well away from any mass threshold $\widetilde T \not= 
\widetilde{m}_A$ (so that the ``spring constant'' term is relatively 
negligible in Eq.~(\ref{eq3.21})) any initial ``velocity'' 
$\dot{\varphi}_{\rm in} = \chi_{\rm in}$ (satisfying $ \chi_{\rm in}
\leq \sqrt{3} H_{* {\rm in}}$ from Eq.~(\ref{eq3.9b}) with $k = 0$)
will be  generically quickly (in a few 
$e$-folds) damped away (as said above, in these circumstances,
$\dot{\varphi}$ decreases like $R_*^{- 3}$). Therefore, a physically
most natural requirement is to impose (as we do in this work) that the
initial velocity $\chi_{\rm in}$ {\it vanish} when starting a BBN run at a 
temperature $\widetilde{m}_{\mu} \gg \widetilde T \gg \widetilde{m}_e$. 
[The physical effects due to $\varphi$ would be stable under using
only an inequality $\chi_{\rm in} \ll H_*$, instead of a strict
vanishing of  $\chi_{\rm in}$.] Evidently, there exist mathematical
solutions to Eq.~(\ref{eq3.21}) where $ \dot{\varphi}/H_*$ is not small
before BBN. In view of the argument presented above, and in absence
of a generic mechanism giving rise to large (i.e. very near $\sqrt{3}$)
 initial values of $ \dot{\varphi}/H_*$,
we think that such solutions are physically irrelevant.

On the other hand, the 
initial ``position'' $\varphi_{\rm in}$ is physically unrestricted, at 
least if the curvature parameter $\beta$ is not very much larger than 
unity. [It was shown in Refs.~\cite{DN}, \cite{DP} that when $\beta > 10$, 
the effect of the previous mass thresholds (above the electron one) is very 
efficient in attracting $\varphi$ toward zero, so that in this case one 
would expect $\varphi_{\rm in}$ before BBN to be small.] In this work, we 
shall formally leave $\varphi_{\rm in}$ unrestricted, even when $\beta \sim 
100$, to see what constraints BBN brings on this possibility. As the 
explicit numerical calculations we made start at the temperature 
$\widetilde{T}_0 = 10^{11} \, K \sim 10 \, {\rm Me V}$ which is only 20 
times larger than $\widetilde{m}_e$, we have slightly refined the initial 
conditions by taking into account the attracting effect of the source term 
$\sigma_*$ in Eq.~(\ref{eq3.21}), integrated over the temperatures 
$\widetilde{m}_{\mu} > \widetilde T \gg \widetilde{T}_0$. Using the 
asymptotic behaviour $\widetilde{\sigma}_e \propto \widetilde{T}^{-2}$ as 
$\widetilde T \gg \widetilde{m}_e$, one finds that, starting from $\varphi 
= \varphi_{\rm in}$, $\chi = \chi_{\rm in} =0$ at $\widetilde{m}_{\mu} > 
\widetilde T \gg \widetilde{T}_0 \gg \widetilde{m}_e$, one ends up with the 
following corrected initial conditions at $\widetilde T = \widetilde{T}_0$:
\begin{mathletters}
\label{eq3.23}
\begin{eqnarray}
&&\varphi (\widetilde{T}_0) \simeq \varphi_{\rm in} - \frac{1}{4} \, 
\alpha_{\rm in} \left( \frac{\widetilde{\sigma}}{\widetilde{\rho}} 
\right)_{\widetilde{T}_0} \, , \label{eq3.23a} \\
&&\chi (\widetilde{T}_0) \simeq - \frac{1}{2} \, H_* \, \alpha_{\rm in} 
\left( \frac{\widetilde{\sigma}}{\widetilde{\rho}} \right)_{\widetilde{T}_0} 
\, , \label{eq3.23b}
\end{eqnarray}
\end{mathletters}
where $\alpha_{\rm in} = \alpha (\varphi_{\rm in})$. This ``edge 
correction'' refinement had, anyway, a negligible effect on our results. 
The important point to remember of this discussion is that the forward 
integration in time leads to a physically straightforward, and well 
motivated, choice for the initial time-derivative of $\varphi$. By 
contrast, any integration backward in time is equivalent to integrating an 
oscillator with {\it negative friction} (time-reverse of (\ref{eq3.21}), 
with $\overline f = -f = - 3 H_*$). Such an integration is unstable and 
may generically lead to including (without knowing it) a ``runaway'' mode 
in $\varphi$, i.e. a spurious backward-growing mode (equivalent to setting 
a large, non zero value of $\chi_{\rm in}$ at $\widetilde T \gg 
\widetilde{m}_e$). We think that this incorrect inclusion of spurious 
runaway modes affects the methods used in Refs.~\cite{SKW} and \cite{SA}, 
and may have invalidated some of their results.

\section{Results of numerical integrations and their comparison with 
observational data}

\subsection{Primordial abundances: observations}

The uncertainties in the ``observed'' primordial abundances of the light 
elements (Deuterium, Helium 3, Helium 4, Lithium 7, $\ldots$) are large and 
frought with systematic errors which are difficult to assess. See 
Refs.~\cite{Reeves}, \cite{CST95} for recent overviews of the observational 
situation. As the aim of the present paper is to set conservative limits on 
possible deviations from the ``standard'' (Einstein-gravity) BBN we wish to 
work with the most extreme (hopefully ``$3 \sigma$''-type) ranges for the 
possible ``observed'' primordial abundances. We shall take
\begin{mathletters}
\label{eq4.1}
\begin{eqnarray}
&&0.21 < Y (^4{\rm He}) < 0.255 \, , \label{eq4.1a} \\
&&1.5 \times 10^{-5} < {\rm D} / {\rm H} < 2.4 \times 10^{-4} \, , 
\label{eq4.1b} \\
&&0.7 \times 10^{-10} < {^7{\rm Li}} / {\rm H} < 1 \times 10^{-9} \, . 
\label{eq4.1d}
\end{eqnarray}
\end{mathletters}
Let us comment briefly on our choices. The minimum value $Y_{\rm min} = 
0.21$ of Eq.~(\ref{eq4.1a}) is the one admitted (as extreme possibility) by 
Refs.~\cite{Reeves} and \cite{CST95}, while $Y_{\rm max} = 0.255$ is an 
extreme maximum value suggested by Ref.~\cite{SG95}. The minimum value for 
${\rm D} / {\rm H}$, Eq.~(\ref{eq4.1b}), is from \cite{CST95}, while the 
maximum one was 
derived by assuming that the ``high'' Deuterium value (with its one-sigma 
error bar) reported by Ref.~\cite{RH96} might be correct. We are aware of 
the observational results, and arguments, of Tytler and collaborators 
converging toward a ``low'' Deuterium value, say \cite{BT98},
\begin{equation}
{\rm D}/{\rm H} = (3.4 \pm 0.3) \times 10^{-5} \, , \label{eq4.2}
\end{equation}
but we wish to be very conservative to draw secure limits on tensor-scalar 
gravity. We shall, however, comment below on the effect that a more precise 
knowledge of the primordial Deuterium abundance, of the type of 
Eq.~(\ref{eq4.2}), would have on our results. The minimum value for Lithium 7 
is taken from Ref.~\cite{CST95}, while the maximum one is the extreme value 
chosen in Ref.~\cite{Reeves}. Finally, we have also computed the BBN yield of 
Helium 3. We used an extreme range on the (observationally better controlled) 
sum $({\rm D} + {^3{\rm He}}) / {\rm H}$. However, we found that the 
corresponding limits do not give constraints going beyond those deduced from 
the BBN yields of Helium 4, Deuterium and Lithium.

\subsection{Small $\beta$ case (constant speed up factor)}

Let us first consider the case where the curvature parameter $\beta$ of the 
quadratic coupling function (\ref{eq2.10}) is small, say $\beta \leq 0.2$. 
In the limit $\beta \rightarrow 0$, $\alpha_{\rm quad} (\varphi) = \beta \, 
\varphi \rightarrow 0$, and the evolution equation (\ref{eq3.21}) for 
$\varphi$ becomes simply $\ddot{\varphi} + 3 H_* \, \dot{\varphi} = 0$. The 
generic solution $\varphi$ of this equation tends to an arbitrary constant 
$\varphi = C$ as time evolves in the (normal) positive direction. [The 
other solution is the decreasing mode
 $\dot{\varphi}_{\rm dec} = D / R^3_*$ which 
is quickly damped away.] It was shown in Ref.~\cite{DN} that such a 
behaviour, $\varphi \simeq$ constant, is a good approximation to what 
happens during the $e^+ e^-$ annihilation era {\it when $\beta$ is small}. 
Indeed, Eqs.~(4.20), (4.21) of Ref.~\cite{DN} show that the {\it 
fractional} change of $\varphi$ across the $e^+ e^-$ threshold is of order 
$\Delta_e \, \varphi / \varphi \sim - k_e \, \beta \simeq - 0.16 \, \beta$, 
where the smallish coefficient $k_e$ is proportional to the ratio of the 
number of helicity states of an electron to the total number of 
relativistic degrees of freedom when $\widetilde T > \widetilde{m}_e$. If 
we limit ourselves to values of $\beta \leq 0.2$, the fractional change of 
$\varphi$ will be less than 3\% and can be neglected. This means that, when 
$\beta \leq 0.2$, the effect of tensor-scalar gravity on BBN is simply 
obtained by replacing $\varphi$ by a constant $\varphi_{\rm in}$ and $\chi 
= \dot{\varphi}$ by 0 in Eqs.~(\ref{eq3.8}) and (\ref{eq3.11}) giving the 
physical expansion rate. This yields
\begin{equation}
\widetilde{H}_{({\rm small} \, \beta)} = \sqrt{\frac{8\pi \, G_*}{3} \, A^2 
(\varphi_{\rm in}) \, \widetilde{\rho}} \, . \label{eq4.3}
\end{equation}
By comparison, the standard value of the expansion rate in general 
relativity would read
\begin{equation}
\widetilde{H}^{\rm standard} = \sqrt{\frac{8\pi}{3} \, \widetilde{G}_N \, 
\widetilde{\rho}} \, , \label{eq4.4}
\end{equation}
where $\widetilde{G}_N$ is the presently observed value of Newton's 
constant (as measured in Cavendish-type experiments). The value of this 
observable quantity in tensor-scalar gravity is
\begin{equation}
\widetilde{G}_N = \widetilde{G}_0 = G_* \, A^2 (\varphi_0) \, [1 + 
\alpha_0^2] \, . \label{eq4.5}
\end{equation}
Therefore, the small-$\beta$ limit of the tensor-scalar expansion rate is 
obtained from the standard value by the following rescaling of the standard 
gravitational constant \cite{DN}
\begin{equation}
g \equiv \frac{G_* \, A^2 (\varphi_{\rm in})}{\widetilde{G}_0} = \frac{A^2 
(\varphi_{\rm in})}{A^2 (\varphi_0) \, [1 + \alpha_0^2]} \, . \label{eq4.6}
\end{equation}
Equivalently, one could say that the expansion rate (\ref{eq4.3}) differs 
from its standard counterpart by a constant ``speed up'' factor $\xi \equiv 
\widetilde H / \widetilde{H}^{\rm standard} = \sqrt{g}$. In our quadratic 
model (\ref{eq2.10}), the natural logarithm of $g$ reads, in terms of 
$\alpha_{\rm in} = \alpha (\varphi_{\rm in}) = \beta \, \varphi_{\rm in}$,
\begin{equation}
\ln g = \frac{1}{\beta} \, (\alpha_{\rm in}^2 - \alpha_0^2) - \ln (1 + 
\alpha_0^2) \simeq \frac{1}{\beta} \, (\alpha_{\rm in}^2 - \alpha_0^2) - 
\alpha_0^2 \, . \label{eq4.7}
\end{equation}
Here, we used the fact that $\alpha_0^2 \ll 1$, as known observationally, 
and also consistently deduced below from BBN data. To express $\ln g$ in 
terms of $\alpha_0$ and $\beta$ we need to relate $\alpha_0$ and 
$\alpha_{\rm in}$. In our present approximation, $\alpha_{\rm in} = \beta 
\, \varphi_{\rm in} \simeq \alpha_{\rm out} = \beta \, \varphi_{\rm out}$, 
where $\varphi_{\rm out}$ is the value of $\varphi$ at the end of 
nucleosynthesis. The link between $\varphi_{\rm out}$ and the present value 
$ \varphi_0$ of the scalar field is obtained by integrating 
Eq.~(\ref{eq3.21}) (with initial conditions $\varphi = \varphi_{\rm out}$, 
$\dot{\varphi} = 0$) from the end of nucleosynthesis up till the present 
time. The evolution of $\varphi$ during this time span comes from its 
coupling to the total matter (baryons plus cold dark matter). The 
corresponding scalar source $\sigma_{*m} = \rho_{*m} - 3 p_{*m} \simeq 
\rho_{*m}$ has a small effect on the evolution of $\varphi$ during the 
radiation era but becomes very important during the subsequent 
matter-dominated era. It was very generally shown in Ref.~\cite{DN} that 
the $t_*$-time evolution equation (\ref{eq3.12}) for $\varphi$ (which must 
{\it a priori} be solved together with the evolution equation for the scale 
factor $R_*$) can be conveniently rewritten as a decoupled equation for the 
evolution of $\varphi$ with respect to the parameter
\begin{equation}
p = \int H_* \, dt_* = \ln R_* + {\rm cst} \, , \label{eq4.8}
\end{equation}
which measures the number of $e$-folds in the Einstein frame. The latter 
$p$-time equation was approximately solved in Ref.~\cite{DN} for the 
present case of the transition between the radiation era and the matter 
era. For simplicity, and because it is the theoretically most plausible
value, we assume here and below that the spatial curvature $k = 0$. [ The
effect of nonzero curvature has been investigated in Ref.~\cite{DN}.]
 The result for small $\beta$ reads
\begin{equation}
\frac{\varphi_0}{\varphi_{\rm out}} = \frac{\alpha_0}{\alpha_{\rm out}} 
\simeq \frac{1+r}{2r} \, \exp \left[ - \frac{3}{4} \, (1-r) \, p_0 \right] 
\simeq e^{-\beta p_0} \, . \label{eq4.9}
\end{equation}
Here, $r \equiv \sqrt{1-8\beta / 3} = 1 - 4\beta / 3 + {\cal O} (\beta^2)$, 
and
\begin{equation}
p_0 \equiv \ln (R_{*0} / R_*^{\rm equivalence}) \, , \label{eq4.10}
\end{equation}
denotes the number of Einstein-frame $e$-folds separating the present moment 
from the time 
of equivalence between matter and radiation: $1 = (\rho_{*m} / \rho_{* \, 
{\rm rad}})^{\rm equivalence} = (\widetilde{\rho}_m / \widetilde{\rho}_{\rm 
rad})^{\rm equivalence}$. The corresponding {\it physical} redshift is
\begin{equation}
\widetilde{Z}_0 = \frac{\widetilde{R}_0}{\widetilde{R}^{\rm equivalence}} = 
\frac{\widetilde{\rho}_m^0}{\widetilde{\rho}_{\rm rad}^0 } = 2.404 \times 
10^4 \, \Omega_m \, h^2 \, . \label{eq4.11}
\end{equation}
Here, $\Omega_m \equiv \widetilde{\rho}_m^0 / \widetilde{\rho}_c^0$ is the 
present ratio of the total matter density to the (physical-units) ``closure 
density'' $\widetilde{\rho}_c^0 = 3 \, \widetilde{H}_0^2 / (8 \pi \, 
\widetilde{G}_0)$ and $h \equiv \widetilde {H}_0 / (100 \, {\rm km s}^{-1} 
\, {\rm Mpc}^{-1})$ is the reduced Hubble parameter. Note that the product
\begin{equation}
\Omega_m \, h^2 = \frac{\widetilde{\rho}_m^0}{1.879 \times 10^{-29} \, {\rm 
g} \, {\rm cm}^{-3}} \, , \label{eq4.12}
\end{equation}
is independent from the value of $\widetilde{H}_0$ and depends only on the 
actual value of the total matter density. Finally, using the link 
(\ref{eq3.6b}) between $\widetilde R$ and $R_*$, the number of 
Einstein-frame $e$-folds (\ref{eq4.10}) reads \cite{DN}
\begin{equation}
p_0 = \ln \widetilde{Z}_0 + a (\varphi_{\rm out}) - a (\varphi_0) \, , 
\label{eq4.13}
\end{equation}
which gives numerically
\begin{eqnarray}
p_0 &=& 10.09 + \ln (\Omega_m \, h^2) + a (\varphi_{\rm out}) - a (\varphi_0) 
\nonumber \\
&=& 8.19 + \ln (\Omega_m \, h^2 / 0.15) + a (\varphi_{\rm out}) - a 
(\varphi_0) \, . \label{eq4.14}
\end{eqnarray}
Here, we introduced as fiducial value for the matter density the value 
$(\Omega_m \, h^2)^{\rm fiducial} = 0.15$ which corresponds roughly to the 
currently favoured values $\Omega_m \sim 0.3$, $h \simeq 0.7$ (it could 
also correspond to $\Omega_m \simeq 1$, $h \simeq 0.4$).

\begin{figure}
\hspace{-1cm}
\vspace{+4cm}
\centerline{\epsfig{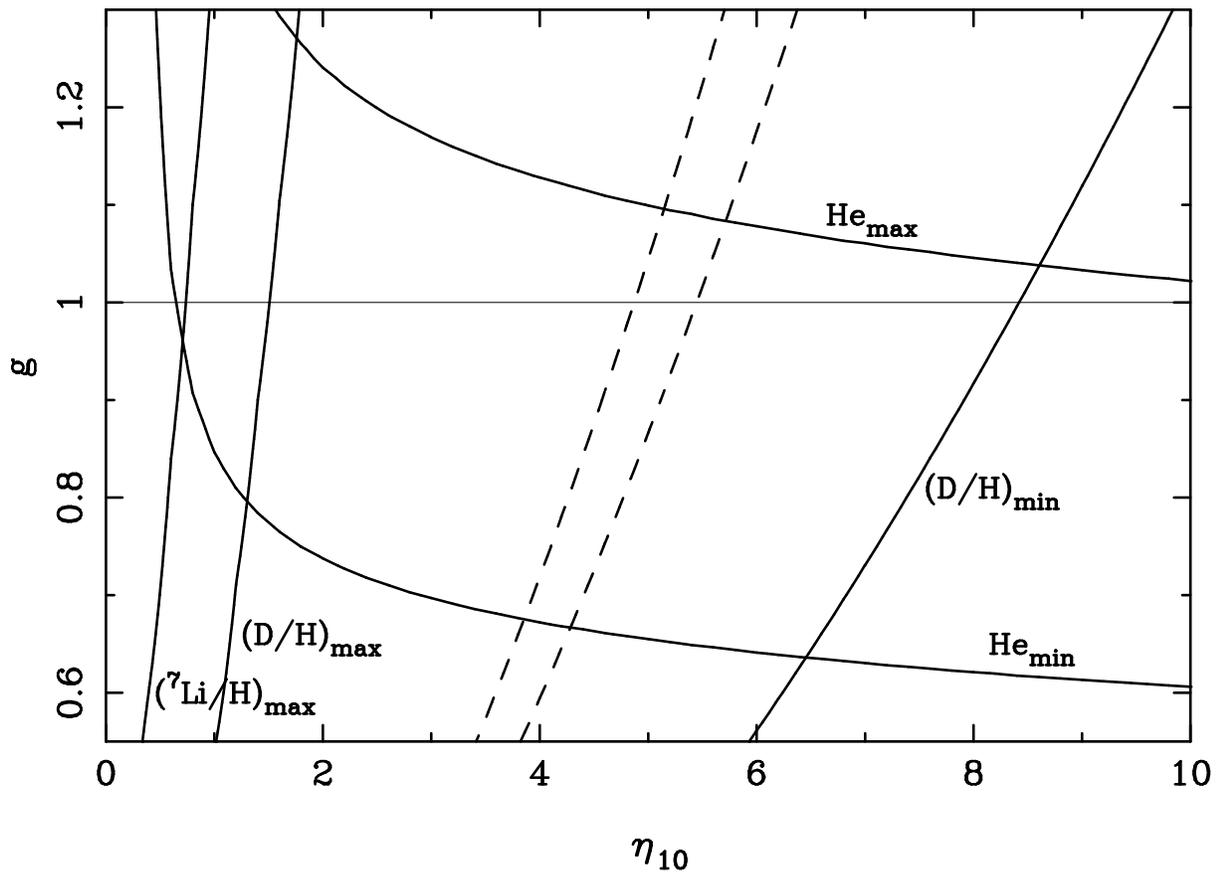}}
\caption{Level contours of BBN yields as functions of the baryon-to-photon 
ratio $\eta_{10} = 10^{10} \, \eta = 10^{10} \, \widetilde{n}_b / 
\widetilde{n}_{\gamma}$ and of a (squared) speed-up factor $g = \xi^2 = 
\widetilde{G}^{\rm effective} / \widetilde{G}_{\rm Newton}$. The solid lines 
correspond to the conservative observational bounds of Eqs.~(\ref{eq4.1}). 
The allowed region is the inside of the curved parallelogram defined by the 
${\rm He}$ and ${\rm D} / {\rm H}$ lines. The dashed lines, given for 
illustration, correspond to the Deuterium $1\sigma$ range of 
Eq.~(\ref{eq4.2}).}
\label{fig1}
\end{figure}

Using the link (\ref{eq4.9}) we can express $\alpha_{\rm in}$, in 
Eq.~(\ref{eq4.7}), in terms of $\alpha_0 : \alpha_{\rm in} \simeq 
\alpha_{\rm out} \simeq e^{+\beta \, p_0} \, \alpha_0$. This gives
\begin{equation}
\ln g \simeq \frac{\alpha_0^2}{\beta} \, [e^{2\beta \, p_0} - 1 - \beta ] 
\, . \label{eq4.15}
\end{equation}
Note that, though we are considering here smallish values of $\beta \sim 
0.1$, we are not entitled to expanding the exponential in 
Eq.~(\ref{eq4.15}). Indeed, we see from Eq.~(\ref{eq4.14}) that $2p_0 \sim 
16$ is rather large. When $\beta$ is very small (with $2 p_0 \, \beta \ll 
1$), the expansion of the R.H.S. of Eq.~(\ref{eq4.15}) gives a finite limit
\begin{equation}
(\ln g)_{\beta \rightarrow 0} = (2 p_0 - 1) \, \alpha_0^2 \, . 
\label{eq4.16}
\end{equation}
It is easily checked that the R.H.S. of Eq.~(\ref{eq4.15}) is positive 
(because $p_0 > 1/2$). Therefore, we conclude that, in the limit of small 
$\beta$, tensor-scalar gravity leads to speeding up the expansion rate: 
$\xi = \sqrt g > 1$. It is, however, interesting to leave first 
unrestricted the sign of $\ln g$ and to study the effect on BBN of 
replacing $\widetilde{G}_N \rightarrow g \, \widetilde{G}_N$ for both signs 
of $\ln g$. We have run BBN codes with varying values of $g$ and $\eta = 
\widetilde{n}_b / \widetilde{n}_{\gamma}$,
 and computed the resulting abundances of light elements, 
comparing them to the extreme observational ranges given in 
Eqs.~(\ref{eq4.1}). The results of this comparison are given in 
Fig.~\ref{fig1} which represents the allowed regions in the $(\eta , g)$ 
plane. The total phenomenologically allowed range for $g$ would be $g_{\rm 
min} < g < g_{\rm max}$ with
\begin{equation}
g_{\rm min} \simeq 0.64 \ , \ g_{\rm max} \simeq 1.27 \, . 
\label{eq4.17}
\end{equation}
In our case $g$ is restricted to be larger than one and we have the firm 
constraint $1 \leq g < g_{\rm max}$. Note that the maximum allowed value of 
$g$ is constrained by the upper bound on the deuterium, combined with the 
upper bound on Helium 4. To illustrate the effect of choosing a much smaller 
range for deuterium, we have also plotted on Fig.~\ref{fig1} (in dashed 
lines) the limits on $g$ obtained by using as confidence interval the 
$(1\sigma)$ one suggested by the recent results of Tytler and 
collaborators \cite{BT98}, Eq.~(\ref{eq4.2}). This would give the stronger 
constraint $g_{\rm max} \simeq 1.1$.

Note also that the shapes of the curves 
bounding the maximally allowed region are such that, when $g$ is only allowed 
to 
vary above one, the corresponding allowed interval for the values of the 
baryon to photon ratio $\eta_{10} \equiv 10^{10} \times \eta$ does not 
change significantly: roughly $1.5 < \eta_{10} < 8.5$. Therefore, when 
$\beta < 0.2$, the addition of a scalar component to gravity does not change 
the upper bound on $\Omega_b \, h^2$ deduced from BBN. The case $\beta > 
0.2$ will be considered below. 

Let us now turn to the constraints on the 
parameters $\alpha_0$ and $\beta$ of tensor-scalar gravity following from 
BBN. Using the result (\ref{eq4.15}), the constraint $1 \leq g < g_{\rm max}$ 
translates in the following bound on $\alpha_0^2$
\begin{equation}
\alpha_0^2 < \frac{\beta}{e^{2\beta \, p_0} -1 -\beta} \, \ln g_{\rm max} 
\, . \label{eq4.18}
\end{equation}
In particular, the limit (\ref{eq4.16}) when $\beta \ll 1/(2 p_0)$ yields
\begin{equation}
(\alpha_0^2)_{\beta \rightarrow 0} < \frac{1}{2 p_0 -1} \, \ln g_{\rm max} 
\, , \label{eq4.19}
\end{equation}
which is numerically of order $(\alpha_0^2)_{\beta \rightarrow 0} \lesssim 
0.015$. Note that, given a value of $\Omega_m \, h^2$, the precise value 
of $p_0$ defined by Eq.~(\ref{eq4.14}) must be obtained by iteration 
because the small (but non negligible) additional term
\begin{equation}
a (\varphi_{\rm out}) - a (\varphi_0) = \frac{\alpha_{\rm out}^2 - 
\alpha_0^2}{2\beta} = \frac{e^{2\beta \, p_0} - 1}{2\beta} \, \alpha_0^2 \, 
, \label{eq4.20}
\end{equation}
depends both on $\alpha_0$ and on $p_0$. However, in our present 
approximation of small $\beta$, we can replace (\ref{eq4.18}) in 
(\ref{eq4.20}) to get
\begin{equation}
(a_{\rm out} - a_0)_{{\rm small} \, \beta}^{\rm maximum} = \frac{e^{2\beta \, 
p_0} - 1}{2(e^{2\beta \, p_0} - 1 - \beta)} \, \ln g_{\rm max} \simeq 
\frac{1}{2} \, \ln g_{\rm max} \, . \label{eq4.20new}
\end{equation}
We then deduce from this result that the value of $p_0$ which can be used for 
estimating the maximum BBN-allowed $\alpha_0^2$ is
\begin{eqnarray}
(p_0)_{{\rm small} \, \beta} &\simeq& 8.19 + \ln (\Omega_m \, h^2 / 0.15) + 
\frac{1}{2} \, \ln \, g_{\rm max} \nonumber \\
&\simeq& 8.31 + \ln (\Omega_m \, h^2 / 0.15) \, . \label{eq4.21}
\end{eqnarray}

\subsection{$\beta > 0.2$ case}

Let us now consider the complementary case where $\beta > 0.2$. In this 
case, the total attraction due to the matter era is quite large. Indeed, 
the ratio $\alpha_0^2 / \alpha_{\rm out}^2$ decreases fast with $\beta$, 
and even for $\beta = 0.2$, one finds from Eq.~(\ref{eq4.9}) $\alpha_0^2 / 
\alpha_{\rm out}^2 \simeq e^{-2\beta \, p_0} \sim 0.04$, where we used $p_0 
\sim 8$. Therefore, when $\beta \geq 0.2$ we can, as was mentioned in 
Section 3, neglect the difference between $G_*$ and $\widetilde{G}_0 = G_* 
\, A^2 (\varphi_0) [1 + \alpha_0^2]$. This allows one to run (in the 
forward time direction) the tensor-scalar-modified BBN code described in 
Section 3, with the numerical value $G_* \simeq 6.672 \times 10^{-8} \, 
{\rm cm}^3 \, {\rm g}^{-1} \, {\rm s}^{-2}$. Let us recall that the input 
parameters 
which must be chosen in each run are: $\beta$, the baryon-to-photon ratio 
$\eta_{10} = 10^{10} \times \eta$, and the value $\varphi_{\rm in}$ of the 
scalar field before freeze-out and $e^+ e^-$ annihilation. Instead of 
working with $\varphi_{\rm in}$, it is physically more appropriate to work 
with the corresponding value of the (logarithmic) coupling function $a_{\rm 
in} \equiv a (\varphi_{\rm in}) = \frac{1}{2} \, \beta \, \varphi_{\rm 
in}^2$. For each value of the triplet $(\beta , \eta , a_{\rm in})$, the 
code then computes the abundances of light elements produced by the BBN, 
and the value $\varphi_{\rm out}$ of the scalar field after the end of
nucleosynthesis (together with the 
corresponding $a_{\rm out} = a (\varphi_{\rm out})$ and $\alpha_{\rm out} = 
\alpha (\varphi_{\rm out})$). Let us recall that for temperatures well above 
and well below the electron-annihilation threshold 
$\widetilde T \sim \widetilde{m}_e$, the 
source term $\widetilde{\sigma}_e$ on the R.H.S. of the evolution equation 
(\ref{eq3.13b}) for $\varphi$ becomes negligible, while the friction term 
remains, so that $\varphi$ stops evolving. [This approximation is correct 
as long as $\widetilde T \ll \widetilde{m}_{\mu}$ and $\widetilde T \gg 
\widetilde{T}_{\rm equivalence}$.] On the other hand, when $\widetilde T$ 
is between $5 \, \widetilde{m}_e \sim 2.5 \, {\rm Me V}$ and $0.2 \, 
\widetilde{m}_e \sim 0.1 \, {\rm Me V}$, which is a crucial period 
during which weak interactions freeze (the $(n/p)$ ratio freezes out around 
$\widetilde T \sim 0.8 \, {\rm Me V}$), and nuclear reactions start to 
build up light elements, the source term $\widetilde{\sigma}_e$ is 
numerically important and causes $\varphi$ to oscillate around $\varphi = 
0$. As discussed in detail in Ref.~\cite{DP} these oscillations are very 
vigorous when $\beta \gg 1$. Therefore, we have a physically non trivial 
situation where the physical cosmological expansion rate $\widetilde H$ 
(given in terms of $\varphi$ by Eqs.~(\ref{eq3.8}) and (\ref{eq3.11})) 
varies in a complicated manner precisely during the crucial stages of BBN. 
This situation cannot be analytically approximated, and must be tackled 
numerically. We can visualize the results of our numerical simulations by 
representing, for each value of $\beta$, the contour levels of the light 
element abundances in the $(\eta , a_{\rm in})$ plane. The topology of 
these contour levels depend on how large $\beta$ is.

\begin{figure}
\hspace{-1cm}
\vspace{+4cm}
\centerline{\epsfig{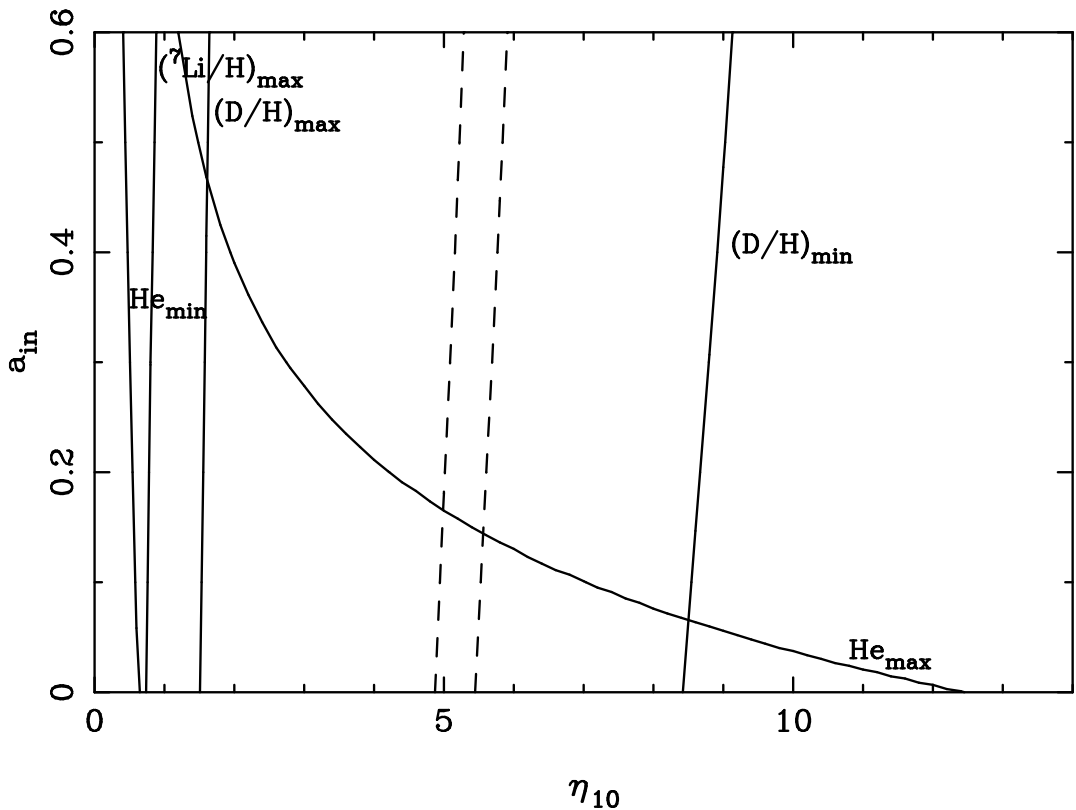}}
\caption{Level contours of BBN yields when $\beta = 10$ as functions of the 
baryon-to-photon ratio $\eta_{10}$ and of the initial value of the scalar 
coupling function $a_{\rm in} = a (\varphi_{\rm in})$. The lines are defined 
as in Fig.~\protect{\ref{fig1}}.
 The allowed region for the conservative bounds is the 
truncated quasi-triangle defined by ${\rm He}_{\rm max}$, $({\rm D} / {\rm 
H})_{\rm max}$, $({\rm D} / {\rm H})_{\rm min}$ and the horizontal axis.}
\label{fig2}
\end{figure}

When $\beta \lesssim 16$, the contour levels look like continuously 
deformed versions of the small-$\beta$ case, i.e. the case of a constant 
speed up factor $\xi = \sqrt g$ shown above in Fig.~\ref{fig1}. For instance, 
the case $\beta = 10$ is shown in Fig.~\ref{fig2} and should be compared with 
the part of Fig.~\ref{fig1} above the $g=1$ line. Therefore, when $\beta 
\lesssim 16$ the upper bound on the value of $a_{\rm in}$ is given, as in the 
small-$\beta$ case, by the upper left corner of the ``triangle'' defined by 
the $({\rm D} / {\rm H})_{\rm max}$ and the ${\rm He}_{\rm max}$ lines. We 
have checked that there 
is approximate continuity between the small-$\beta$ case and the $\beta > 
0.2$ case: Indeed, by running a BBN code for $\beta = 0.2$ we find a 
maximum allowed value $a_{\rm in}^{\rm max} = 0.125$ (together with a
corresponding $a_{\rm out}^{\rm max} \simeq 0.109$) which 
approximately agrees with $\frac{1}{2} \, \ln g_{\rm max} \simeq 0.12 $,
for the $g_{\rm max} \simeq 1.27$ obtained above in the small-$\beta$ 
approximation. [Here, we used Eq.~(\ref{eq4.7}), in which we neglected the 
fractionally small $\alpha_0^2$ terms.]  
As in the small-$\beta$ case, we find that the shapes of the curves
bounding the allowed region are such that the corresponding allowed interval
for $\eta_{10}$ does not change significantly.

When $\beta \gtrsim 16$, the ``triangle'' of Fig.~\ref{fig2} disappears 
because many 
of the contour lines tend to straighten out and become parallel. This 
situation is illustrated in Fig.~\ref{fig3} for $\beta = 40$. This case 
illustrates two interesting phenomena. First, though the shapes of the
contours have changed a lot, they are still such that the allowed interval
for $\eta_{10}$ does not change significantly: roughly, $1.5<\eta_{10}<8.5$.
 Second, Fig.~\ref{fig3} shows that the 
initial value $a_{\rm in}$, i.e. the initial speed-up factor $\xi_{\rm in} 
= A_{\rm in} / (A_0 \sqrt{1+\alpha_0^2}) \simeq A_{\rm in} = \exp (a_{\rm 
in})$ 
can be extremely large (for instance, the largest value $a_{\rm in} \simeq 
5$ of Fig.~\ref{fig3} corresponds to $\xi_{\rm in} \simeq 150$, and much 
larger 
values are possible) and still lead to compatibility with the observed 
abundances of light elements. This result contradicts the usual expectation 
that the BBN data are tight enough to constrain to a small level any 
deviation from standard gravity around the epoch of BBN.
 Here, we have an example where, just 
before the BBN starts, the tensor-scalar-predicted expansion rate is 
several orders of magnitude larger than the Einstein-predicted one, and 
where the BBN test is still passed.
This paradoxical result raises two questions: (i) How can an 
abnormally large expansion rate, around the (nominal) time of
freeze-out of the neutron to proton ratio, lead to quasi-normal light
element abundances ?; and, (ii) Does this imply that, when $\beta 
\gtrsim 16$, the present value of $\alpha_0^2$ (which measures the 
deviation between general relativity and tensor-scalar gravity) can be 
large? 

 To answer the first question, we plot in Fig.~\ref{fignew} the coupling
function $a(\varphi)$, which is essentially the logarithm of the speed-up
factor, as a function of the inverse (physical) temperature 
$ 1/\widetilde{ T} $ (we use as input values $\beta = 40$, and 
$a_{\rm in} = 2$). We plot on the same Figure the evolution with 
temperature of: the actual neutron to proton ratio in tensor-scalar theory
 (solid line), the  $(n/p)$ ratio in the standard general relativistic
case (dashed line), and the instantaneous equilibrium value for this ratio:
$(n/p)_{\rm eq} = \exp (- Q/\widetilde {T})$, where $ Q \equiv
\widetilde{m}_n - \widetilde{m}_p \simeq 1.293$ MeV (dotted line).
If we define (as is often done) a nominal freeze-out temperature
as the temperature $\widetilde{T}_*$ where the actual $(n/p)$
ratio starts to deviate significantly from $(n/p)_{\rm eq}$ (a sign
that weak interactions become slow compared to the expansion rate),
we see that the value of $\widetilde{T}_*$ in tensor-scalar gravity
is larger than in general relativity, as expected from the fact that
the expansion rate is  much larger ($a(\varphi)$ being initially 
frozen at $a \simeq a_{\rm in} = 2$). If that were all, this would lead
to a significantly larger value for the $(n/p)$ ratio. However, 
Fig.~\ref{fignew} illustrates two physical phenomena that modify this
naive conclusion. First, around $\widetilde {T} \sim 0.8 $ MeV, the source
term $ \widetilde{\sigma}_e$ starts to have a strong effect on $a(\varphi)$:
it makes $a(\varphi)$ drop precipitously toward zero, where it bounces
twice to end up being frozen (when $\widetilde{T} \lesssim 1/20 $ MeV) 
at a very small limiting value $a_{\rm out} \simeq 0.016$. The second 
important phenomenon illustrated by Fig.~\ref{fignew} is the fact that the
freeze-out phenomenon is not something which takes place at a precise 
moment in time, but is an integrated phenomenon. It is clear from the Figure
that the approximate plateau reached by $(n/p)$ (before it drops around 
$1/15$ MeV when nuclear reactions start) is due to the integrated effect
of an evolution which takes place before, during and after the precipitous
fall of $a(\varphi)$ toward small values. It is therefore impossible to 
estimate analytically the final ``plateau'' value of  $(n/p)$  from the
sole knowledge of the nominal freeze-out temperature. 
In particular, we see that $(n/p)$ is
initially larger than the standard GR value, but ends up, in this example,
being somewhat smaller. This Figure illustrates the necessity of resorting
to a full numerical BBN code.

 Let us now address the second question: What does an initially large
value of $a_{\rm in} = a(\varphi_{\rm in})$ imply for the present value
of $\alpha_0^2$ ? Can  $\alpha_0^2$ be large ?
The answer is no, for the following reason: when $\beta$ is large, 
the source $\widetilde{\sigma}_e$, active around $\widetilde T \sim 
\widetilde{m}_e$, is very efficient in attracting $\varphi$ toward zero 
(through a damped-oscillatory evolution). Moreover this attraction effect 
is amplified by nonlinearities so that, for a fixed $\beta$, the post-BBN 
value $a_{\rm out} = a (\varphi_{\rm out})$ does not increase monotonically 
with $a_{\rm in} = a (\varphi_{\rm in})$, but rather decreases for large 
$a_{\rm in}$ values after having reached a maximum when $a_{\rm in} \sim 
1$. This behaviour is illustrated in Fig.~\ref{fig4}. We see from this Figure 
that, 
in the plane of Fig.~\ref{fig3}, $a_{\rm out}$ (considered as a function of 
$a_{\rm 
in}$) reaches the maximum value $a_{\rm out}^{\rm max} \simeq 0.028$ when 
$a_{\rm in} \simeq 0.7$. 

For some particular values of $\beta$ (near 23, 59, 
$\ldots$) the curve giving $a_{\rm out}$ versus $a_{\rm in}$ develops a second 
maximum which becomes higher than the first (like in a first-order phase
transition). This is illustrated in Fig.~\ref{fig5} 
which shows the evolution of the curves between $\beta = 22,23$ and 24, and
between $\beta = 58,59$ and 60. In both cases, the formerly unique maximum 
located near the origin gets superseded, roughly when its location along the
$a_{\rm in}$ axis decreases down to $\sim 0.2$,
by a secondary maximum (which develops abruptly for $\beta \gtrsim 21$, or 
$\beta \gtrsim 56$) located around $a_{\rm in} \sim 1.7$. As $\beta$
further increases, the first maximum gets very small and disappears, 
while the position, along the $a_{\rm in}$ axis, of the second maximum
continuously slides toward the left (so that the ``first maximum'' in
the bottom panel of Fig. 5 is the evolved form of the second maximum 
in the top panel).

In all cases, the 
maximum possible value of $a_{\rm out}$, or better of the difference $a_{\rm 
out} - a_0$ (which matters most when $\beta$ is small), i.e. the maximum 
deviation from general relativity {\it 
after} the end of nucleosynthesis, compatible with BBN data, is quite 
small. This is shown in Fig.~\ref{fig6} which plots the maximum allowed 
$a_{\rm out} - a_0$ in function of $\beta$. When $\beta > 0.2$, the maximum 
value of $a_{\rm out} - a_0 \simeq a_{\rm out}$ is obtained from the results 
of the present sub-section, while for $\beta < 0.2$, $a_{\rm out} - a_0$ is 
obtained from the small-$\beta$ result, Eq.~(\ref{eq4.20new}).

\section{BBN limits on the present deviations from Einstein's theory}

In the previous Section, we have shown what limit BBN data put on $a_{\rm 
out} = a (\varphi_{\rm out})$ as a function of $\beta$. It remains to 
translate this limit on the value of $\varphi$ at the end of 
nucleosynthesis into a limit on the present scalar-coupling parameter 
$\alpha_0 = \alpha (\varphi_0)$. To do this we need to integrate the 
$\varphi$-evolution through the transition between the radiation era and 
the matter era, up till now. As we said above, $\varphi$ satisfies a 
decoupled equation in terms of the parameter $p$ of Eq.~(\ref{eq4.8}) 
\cite{DN}. In the present case of the transition between the radiation era 
and the matter era, it was further shown in Ref.~\cite{DP} that the latter 
$p$-time decoupled evolution equation can be approximately rewritten as the 
following hypergeometric equation
\begin{equation}
x(x+1) \, \partial_x^2 \, \varphi + \left( \frac{5}{2} \, x + 2 \right) \, 
\partial_x  \, \varphi + \frac{3}{2} \, \beta \, \varphi = 0 \, . 
\label{eq5.1}
\end{equation}
Here $x \equiv e^p \equiv R_* (t_*) / R_*^{\rm equivalence}$ denotes the 
(Einstein frame) redshift separating the current time $t_*$ from the time 
of equivalence (when $\rho_{*m} (t_*^{\rm equivalence}) = \rho_{* \, {\rm 
radiation}} (t_*^{\rm equivalence})$). The initial conditions for $\varphi$, 
deep into the radiation era (i.e. for $x \rightarrow 0$), mentioned in 
Section IVB above, select uniquely the solution $\varphi (x) = \varphi_{\rm 
out} \, F [a,b,c;-x]$. Here, $F [a,b,c;z]$ denotes the usual (Gauss) 
hypergeometric series. The values of the parameters are
\begin{equation}
a = \frac{3}{4} - i\omega \ , \ b = \frac{3}{4} + i\omega \ , \ c=2 \ ; \ 
\omega \equiv \sqrt{\frac{3}{2} \left( \beta - \frac{3}{8} \right)} \, . 
\label{eq5.2}
\end{equation}
Finally the ratio between the present value of $\varphi$ and its value at 
the end of nucleosynthesis,
\begin{equation}
\frac{\varphi_0}{\varphi_{\rm out}} = \frac{\alpha_0}{\alpha_{\rm out}} 
\equiv F_m (\beta , Z_{*0}) \label{eq5.3}
\end{equation}
is given by
\begin{equation}
F_m = F [a,b,c;-Z_{*0}] \simeq \frac{\sqrt 2}{\sqrt \pi} \, 2^{2i\omega} \, 
\frac{\Gamma (2i\omega)}{\Gamma \left( 2i\omega + \frac{3}{2} \right)} \, 
e^{-\frac{3}{4} \, p_0} \, e^{i\omega \, p_0} + (i\omega \leftrightarrow 
-i\omega) \, . \label{eq5.4}
\end{equation}
In the last form (made of two terms obtained by changing the sign of 
$\omega$) we have used the asymptotic behaviour, for large argument, of the 
hypergeometric function. Here, $Z_{*0} \equiv e^{p_0}$ denotes the 
Einstein-frame redshift 
separating the present moment from equivalence. Its natural logarithm $p_0 = 
\ln Z_{*0}$ is given by Eq.~(\ref{eq4.14}). The 
result (\ref{eq5.4}) is valid both when $\beta < 3/8$ (in which case 
$\omega$ is pure imaginary, with $i\omega = \sqrt{\frac{3}{2} \left( 
\frac{3}{8} - \beta \right)}$) and when $\beta > 3/8$ ($\omega$ real). It 
is easily checked that when $\beta$ is small the result (\ref{eq5.4}) 
agrees with the more approximate Eq.~(\ref{eq4.9}). Using Eq.~(\ref{eq5.3}) 
we can compute $\alpha_0^2$ in terms of $a_{\rm out} = \frac{1}{2} \, \beta 
\, \varphi_{\rm out}^2 = \alpha_{\rm out}^2 / (2\beta)$
\begin{equation}
\alpha_0^2 = F_m^2 (\beta , Z_{*0}) \, \alpha_{\rm out}^2 = 2 \beta \, F_m^2 
(\beta , Z_{*0}) \, a_{\rm out} \, . \label{eq5.5}
\end{equation}
Using Eq.~(\ref{eq5.5}) we can, for a given value of $Z_{*0}$, translate the 
BBN limit of Fig.~\ref{fig6} on $a_{\rm out} - a_0$ into a limit on the 
present 
scalar-coupling parameter $\alpha_0^2$. As before, when $\beta > 0.2$, 
$a_{\rm out} - a_0 \simeq a_{\rm out}$ and the limit on $\alpha_0^2$ is 
obtained from Eq.~(\ref{eq5.5}), while, when $\beta < 0.2$, one uses the 
small-$\beta$ limit, Eq.~(\ref{eq4.18}). This limit is represented in 
Fig.~\ref{fig7} (solid line)
for our fiducial value $\Omega_m \, h^2 = 0.15$, in Eq.~(\ref{eq4.14}). The 
corresponding limit on the product $\beta \, \alpha_0^2$ is represented in 
Fig.~\ref{fig8}. We recall from Eqs.~(\ref{eq2.9}) that the quantities 
represented 
in Figs.~\ref{fig7} and \ref{fig8} are directly related to the parameters 
measuring 
post-Newtonian deviations from general relativity. As $\alpha_0^2 \ll 1$, 
we have $\overline{\gamma} \equiv \gamma_{\rm Edd} - 1 \simeq -2 \, 
\alpha_0^2$, and $\overline{\beta} \equiv \beta_{\rm Edd} - 1 \simeq 
\frac{1}{2} \, \beta \, \alpha_0^2$.

For comparison let us recall the current direct
experimental limits on $\alpha_0^2$ and $\beta \, \alpha_0^2$. The 
measurement of the Shapiro time delay by the Viking mission 
\cite{reasenberg}, as well as some VLBI measurements \cite{vlbi}, yield the 
$1\sigma$ bound $\vert \overline{\gamma} \vert < 2 \times 10^{-3}$ which 
translates into $\alpha_0^2 < 10^{-3}$. A more stringent limit follows from 
the Lunar Laser Ranging experiment \cite{llr} which yields the $1\sigma$ 
bound $-1.7 \times 10^{-3} < 4 \, \overline{\beta} - \overline{\gamma} < 3 
\times 10^{-4}$. This translates into
\begin{equation}
-8.5 \times 10^{-4} < (1+\beta_0) \, \alpha_0^2 < 1.5 \times 10^{-4} \, 
(1\sigma) \, . \label{eq5.6}
\end{equation}
In the framework of the present paper, the parameter $\beta_0 = \beta$ is 
restricted to be positive. Therefore, the most stringent limit is the 
right inequality in Eq.~(\ref{eq5.6}), i.e.
\begin{equation}
\alpha_0^2 < \frac{1.5 \times 10^{-4}}{\beta + 1} \ , \ \beta \, \alpha_0^2 
< \frac{\beta}{\beta+1} \, 1.5 \times 10^{-4} \, . \label{eq5.7}
\end{equation}
[Strong-field tests put also limits on $\alpha_0^2$ but they are slightly 
less stringent than the above when $\beta$ is positive \cite{DEF}.]
We have  indicated on Figs.~\ref{fig7} and \ref{fig8} the empirical bounds
(\ref{eq5.7}) in dashed lines. 

Figs.~\ref{fig7} and \ref{fig8} show that BBN data put, as soon as $\beta >$ 
0.3, more 
stringent limits on the possible post-Newtonian deviations from general 
relativity than present experimental data. For our fiducial matter density 
$\Omega_m \, h^2 = 0.15$ the level of deviation compatible with BBN 
data is (for $\beta \gtrsim 0.5$) constrained to the level
 $\beta \, \alpha_0^2 \lesssim 10^{-6.5}$. As 
discussed in Ref.~\cite{DN} the attraction factor $F_m^2 (\beta , Z_{*0})$
 in Eq.~(\ref{eq5.5}) scales (for $\beta > 3/8$) like $(\Omega_m \, 
h^2)^{-3/2}$. Note that the corresponding deviation level 
$\alpha_0^2 \lesssim 10^{-6.5} \, \beta^{-1} \,
 (\Omega_m \, h^2 / 0.15)^{-3/2}$ gets 
significantly 
smaller if $\Omega_m \, h^2$ turns out to be of order one. Note finally 
that the attraction factor $F_m$ is dominated by the evolution during the 
matter-dominated era and is only negligibly modified (in view of our other 
approximations) if one takes into account the fact that our recent 
cosmological evolution may have been curvature-dominated \cite{DN} or 
$\Lambda$-dominated.

\section{Conclusions} 

The most natural, and best motivated, alternatives to general relativity are 
the tensor-scalar theories of gravitation. We have considered the class of 
tensor-scalar theories which are dynamically attracted, during their 
cosmological evolution, toward general relativity. The paradigmatic example 
of this class is defined by the quadratic coupling function $a(\varphi) = 
\frac{1}{2} \, \beta \, \varphi^2$ with positive $\beta$. We studied Big Bang 
Nucleosynthesis in this quadratic model as a function of $\beta$, of the 
initial value $\varphi_{\rm in}$ (at temperatures $\sim 10 \, {\rm Me V}$) of 
the scalar field, and of the baryon-to-photon ratio $\eta$. [We assume three 
light neutrinos throughout this work.] We imposed that the theoretically 
predicted BBN yields of light elements (Deuterium, Helium 4 and Lithium 7) be 
compatible with some conservative ranges for the corresponding observed 
primordial abundances, Eqs.~(\ref{eq4.1}).

\begin{figure}
\hspace{-1cm}
\vspace{+4cm}
\centerline{\epsfig{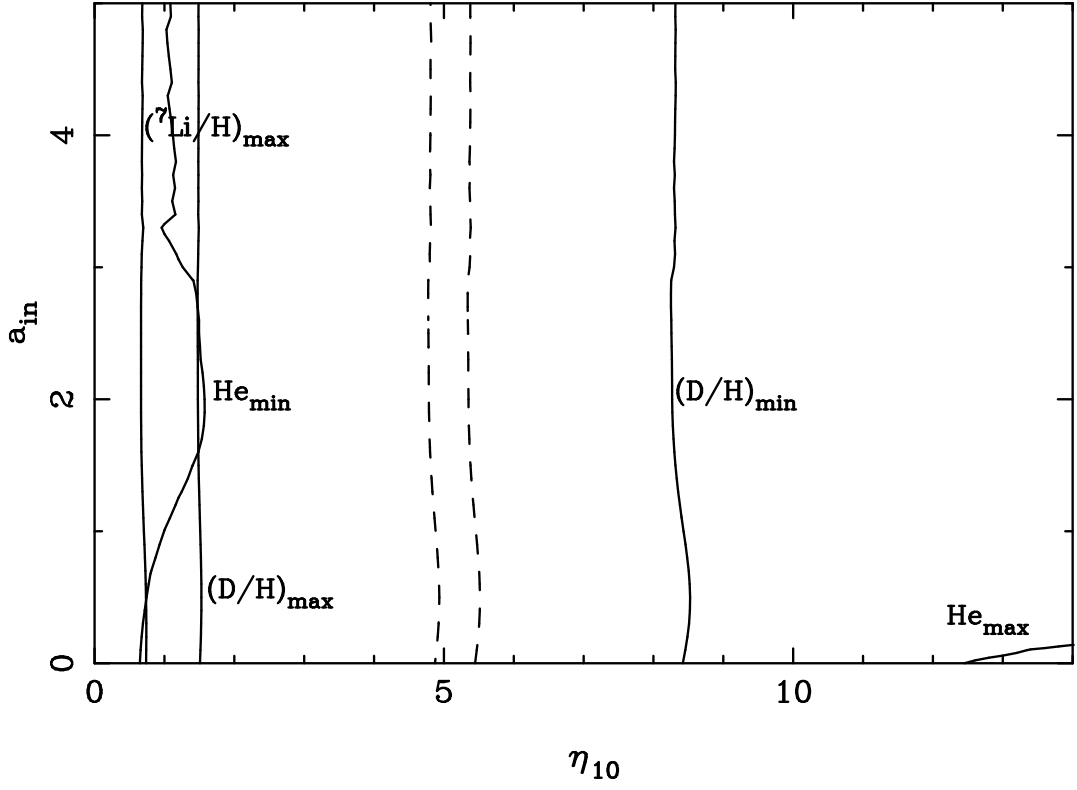}}
\caption{Same as Fig.~\protect{\ref{fig2}}
 for $\beta = 40$. The allowed region has now 
opened into a vertical strip essentially defined by the ${\rm D} / {\rm H}$ 
lines. The wiggles in the ${\rm He}_{\rm min}$ contour signal the presence
of some numerical noise.}
\label{fig3}
\end{figure}

\begin{figure}
\hspace{-3cm}
\centerline{\epsfig{file=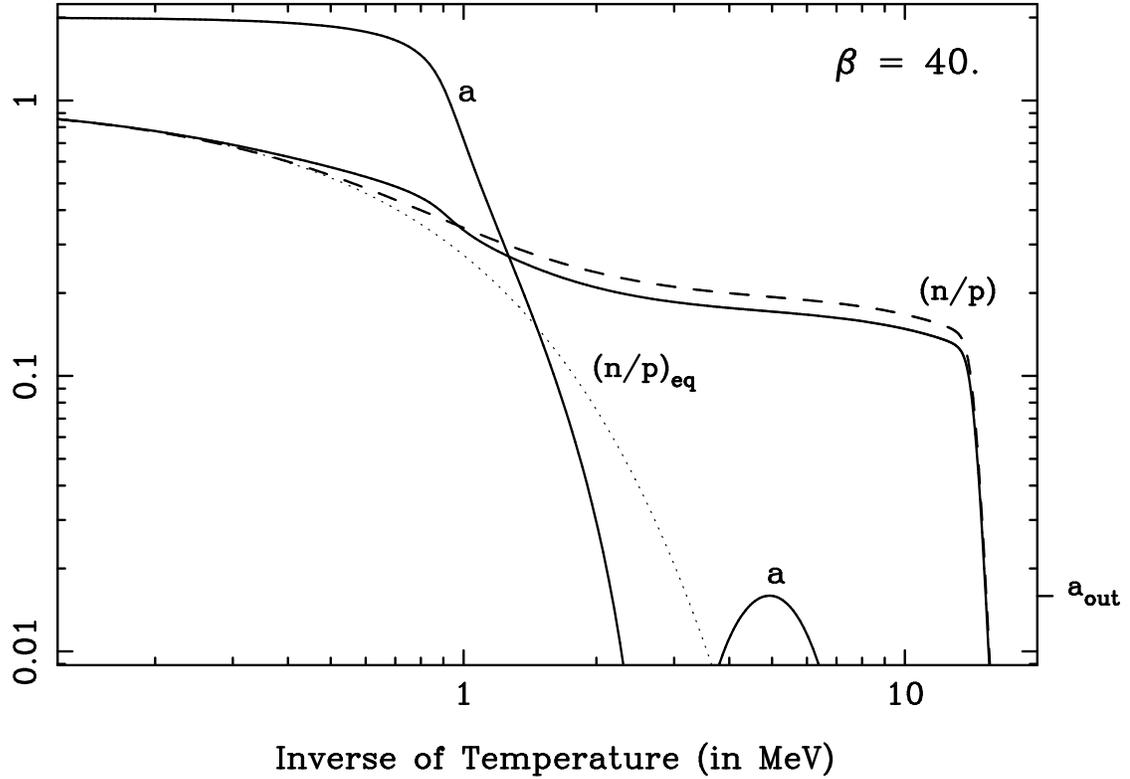,width=1.0\textwidth,angle=-90}}
\caption{ Evolution of the coupling function $a(\varphi)$, and of
the neutron to proton ratio, in function of the inverse of the physical
temperature (for $\beta = 40$ and $a_{\rm in} = 2$). The solid line is
$(n/p)$ in tensor-scalar gravity, the dashed line is $(n/p)$ in general
relativity, while the dotted line is  the equilibrium value of the 
ratio $(n/p)$. $a_{\rm out}$ indicates the asymptotic level reached
by  $a(\varphi)$ for small temperatures.}
\label{fignew}
\end{figure}

\begin{figure}
\hspace{-1cm}
\vspace{+4cm}
\centerline{\epsfig{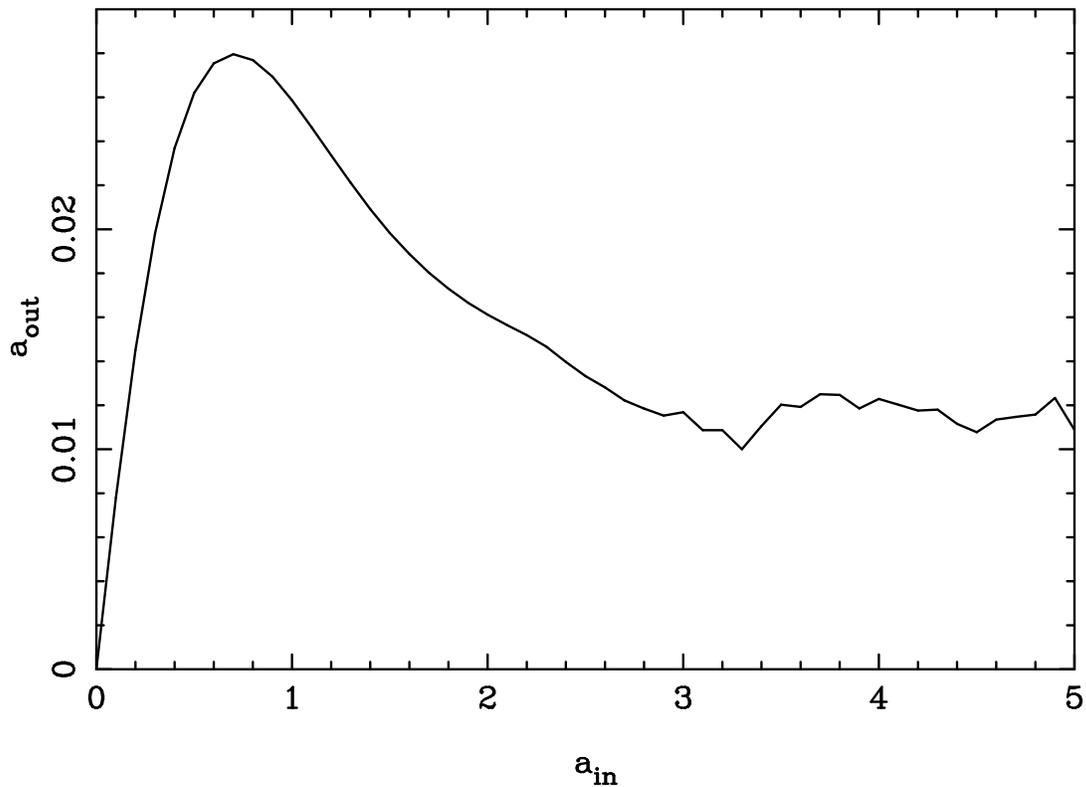}}
\caption{Value of the final scalar coupling function $a_{\rm out} = a 
(\varphi_{\rm out})$, after BBN, in function of its initial value $a_{\rm in} 
= a (\varphi_{\rm in})$ for $\beta = 40$. Because of the presence of a 
maximum in this curve the unlimited allowed region of Fig.\protect{\ref{fig3}}
 still leads to a tight bound on $a_{\rm out}$. The wiggles for 
$a_{\rm in}  > 3$ are due to numerical noise.}
\label{fig4}
\end{figure}

\begin{figure}
\hspace{-1cm}
\vspace{+4cm}
\centerline{\epsfig{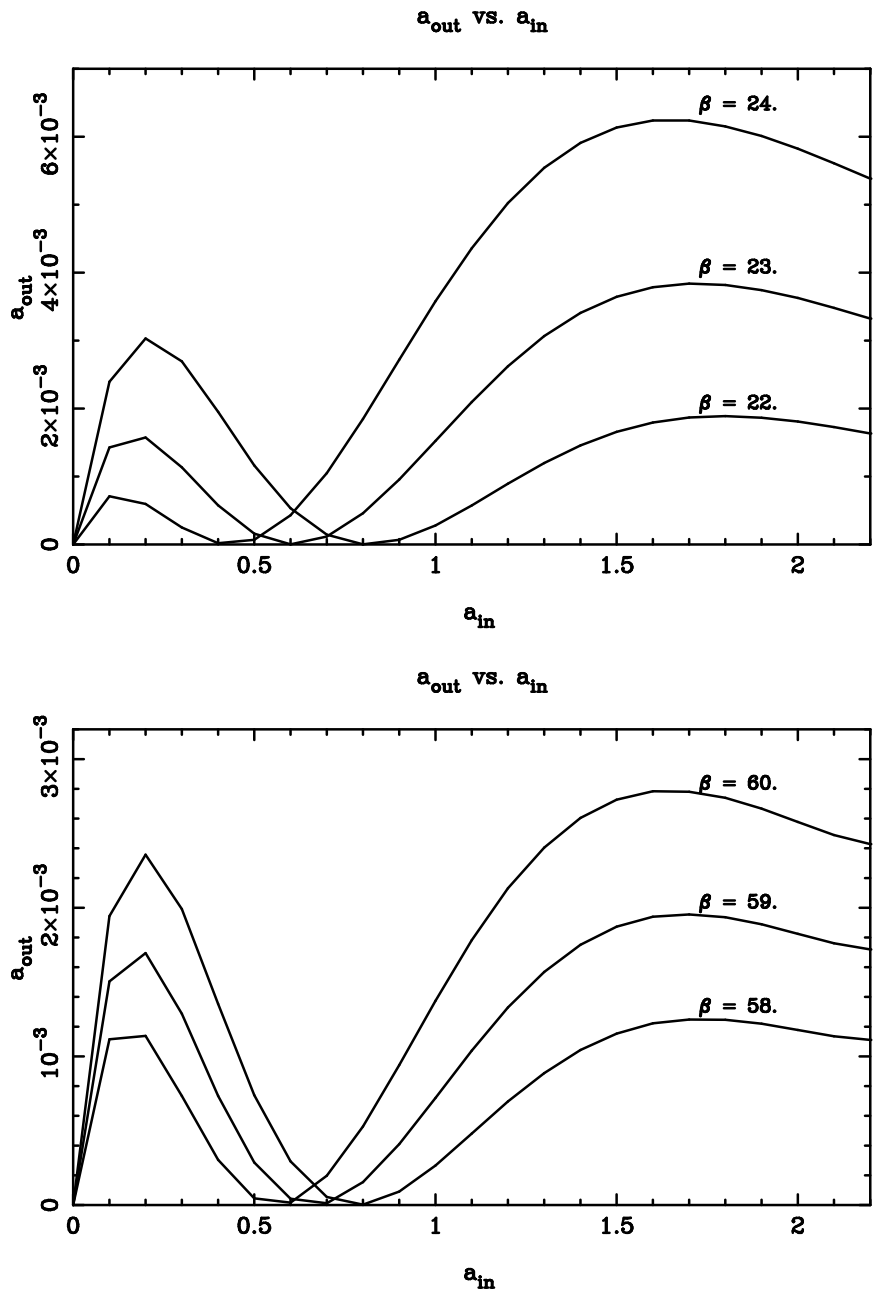}}
\caption{Same as Fig.~\protect{\ref{fig4}}
 for $\beta = 22, 23$ and 24 (top panel),
and for $\beta = 58,59$ and 60 (bottom panel). This figure 
illustrates the competition between two maxima in the curve $a_{\rm out} 
(a_{\rm in})$. For $\beta \geq 23$ and $\beta \geq 59$ it is the second
 maximum which determines the maximum allowed $a_{\rm out}$.}
\label{fig5}
\end{figure}

\begin{figure}
\hspace{-1cm}
\vspace{+4cm}
\centerline{\epsfig{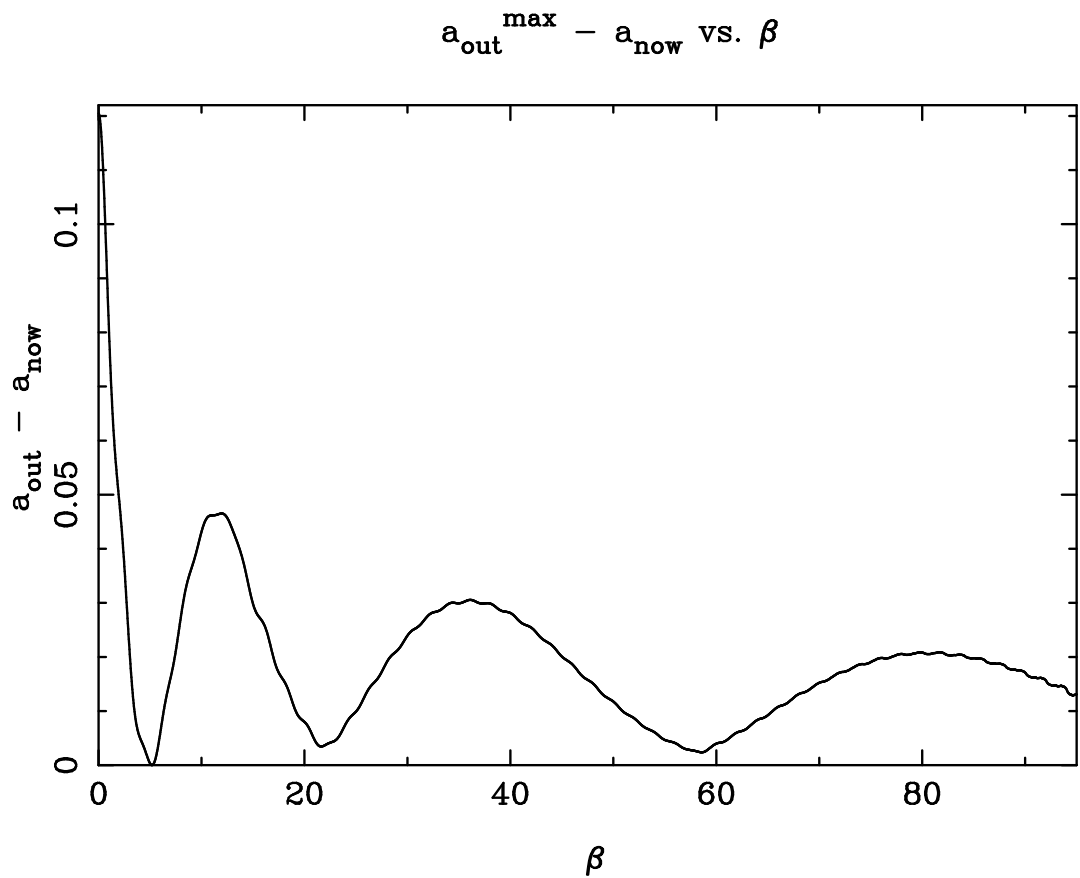}}
\caption{Maximum BBN-allowed value of the difference $a_{\rm out} - a_0$ as a 
function of $\beta$.}
\label{fig6}
\end{figure}

Our first conclusion is that the BBN-inferred upper bound on the cosmological 
baryon density $\Omega_b \, h^2 = 3.66 \times 10^7 \, (\widetilde{T}_0 / 
2.726 \, {\rm K})^3 \, \eta$ is quite robust 
under the addition of a scalar component to gravity. The standard bounds on 
$\eta_{10} \equiv 10^{10} \, \eta$, namely $1.5 < \eta_{10} < 8.5$,
corresponding to $ 0.0055 < \Omega_b \, h^2 < 0.031$,  
 are insignificantly modified even when considering large initial values 
of the scalar field, corresponding to an initial cosmological expansion rate 
much larger than its standard general relativistic value. This is strikingly 
illustrated in Fig.~\ref{fig3}.

\begin{figure}
\hspace{-1cm}
\vspace{+4cm}
\centerline{\epsfig{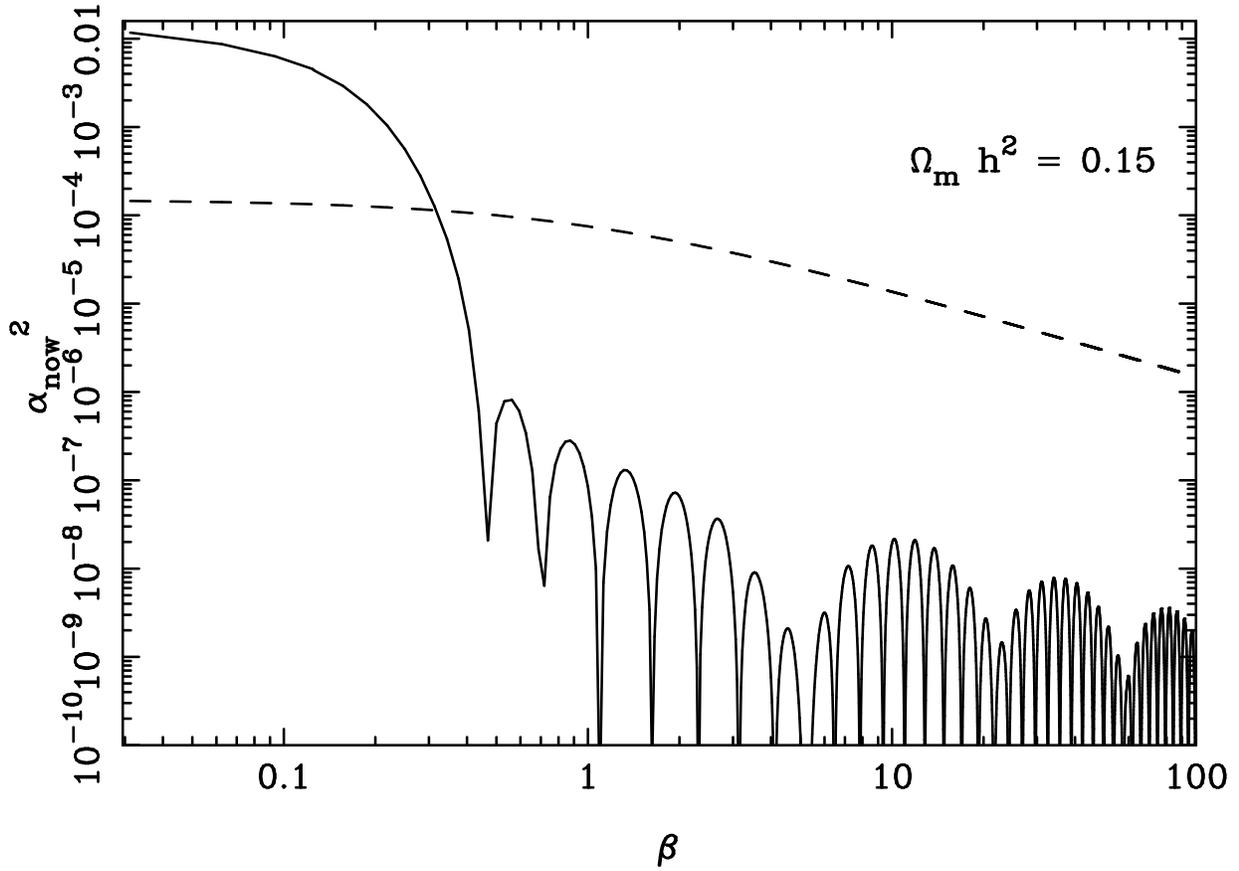}}
\caption{The solid line represents the maximum BBN-allowed value of the 
present scalar-coupling parameter $\alpha_0^2$ as a function of $\beta$. For 
comparison the dashed line represents the most stringent direct empirical 
bound on $\alpha_0^2$.}
\label{fig7}
\end{figure}

\begin{figure}
\hspace{-1cm}
\vspace{+4cm}
\centerline{\epsfig{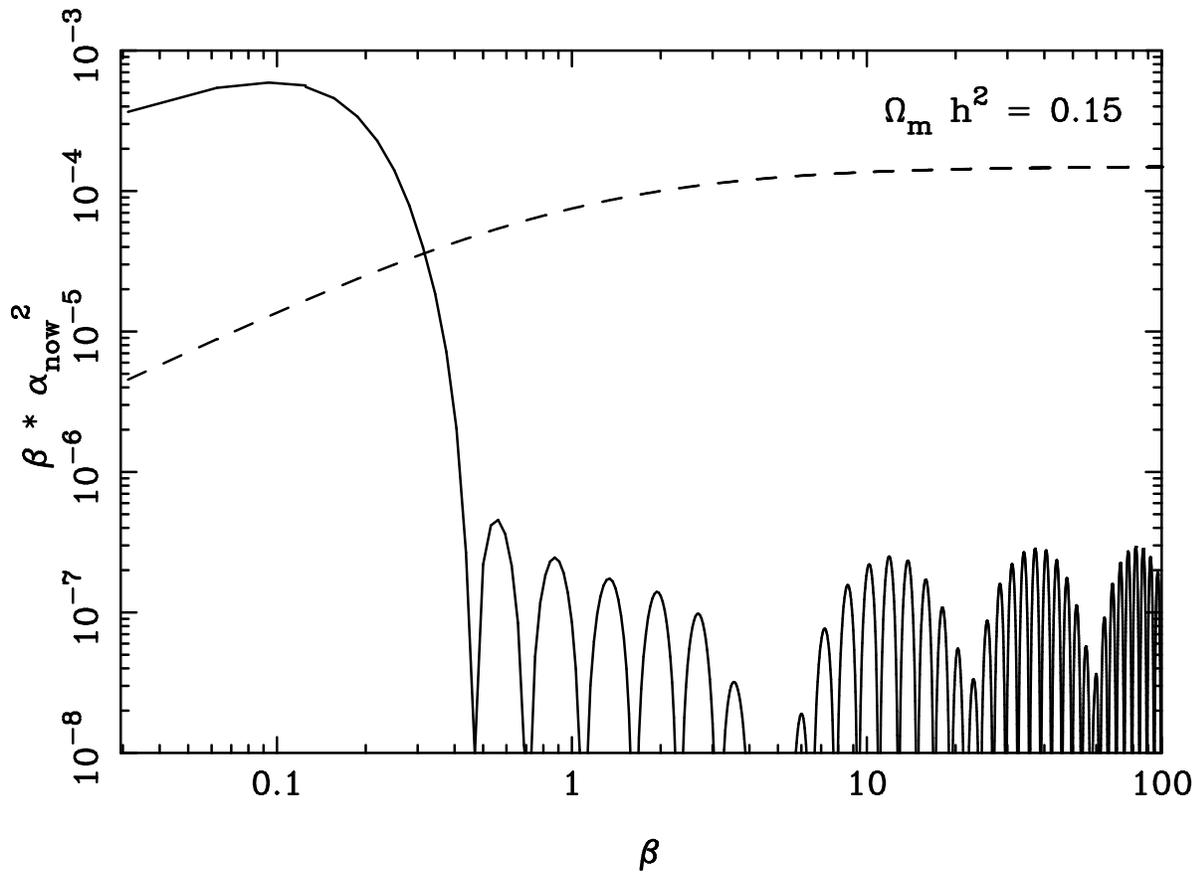}}
\caption{Maximum BBN-allowed value of the product $\beta \, \alpha_0^2$. 
Lines as in Fig.~\protect{\ref{fig7}}.}
\label{fig8}
\end{figure}

Our second conclusion is that, even in the cases where, before BBN, 
tensor-scalar gravity is very different from Einstein's gravity, the 
presently observable deviations from general relativity are constrained by 
BBN to be quite small. The BBN-limits on the two independent weak-field 
``post-Einstein'' parameters $\alpha_0^2$ and $\beta \, \alpha_0^2$ (linked 
to the usual post-Newtonian parameters through $\gamma_{\rm Edd} - 1 \simeq 
-2 \alpha_0^2$, $\beta_{\rm Edd} - 1 \simeq \frac{1}{2} \, \beta \, 
\alpha_0^2$) are exhibited in Fig.~\ref{fig7} and Fig.~\ref{fig8}, in the 
fiducial case where the total matter density is $\Omega_m \, h^2 = 0.15$. The 
BBN-limits in Fig.~\ref{fig7} and \ref{fig8} scale (for $\beta > 3/8$) like 
$(\Omega_m \, h^2 / 0.15)^{-3/2}$. The dashed lines in Figs. 1, 2 and 3
show also that if, in the future, one gets more stringent bounds on the 
Deuterium primordial abundance this will tighten the BBN-limits on
$\alpha_0^2$ for $\beta \lesssim 16$, without affecting them for
$\beta \gtrsim 16$.  As soon as $\beta \gtrsim 0.3$ the 
BBN-limits we obtain on possible deviations from Einstein's theory are much 
stronger than the present observational limits from solar-system or 
binary-pulsar gravitational experiments. They provide motivations for 
experiments which push beyond the present empirical bounds on the basic 
scalar coupling parameter $\alpha_0^2$.

\acknowledgments
We thank Rocky Kolb for kindly providing us with a copy of the
Wagoner-Kawano BBN code, Yvon Biraud for helpful advice on several
numerical issues, and Bob Wagoner for useful comments.

\end{document}